\documentclass[a4paper,11pt,twoside]{article}

\usepackage[margin=32mm,includehead,includefoot]{geometry}
\usepackage{setspace}
\newcommand{\vsp}[1][16pt]{\par \vspace{#1}}

\usepackage{fancyhdr}
\usepackage{titlesec}
\titleformat{\section}
{\centering \Large}{\thesection}{11pt}{}
\titlespacing*{\section}{0pt}{24pt}{8pt}
\titleformat{\subsection}[runin]
{\bfseries}{\normalfont{\thesubsection}}{6pt}{}[.]
\titlespacing*{\subsection}{0pt}{24pt}{6pt}

\usepackage{enumitem}
\setlist{nosep}
\setlist[enumerate,1]{label=\normalfont{(\roman*)}}
\setlist[enumerate,2]{label=\normalfont{(\alph*)}}

\usepackage{tabularray}
\usepackage{graphicx}
\graphicspath{{./}}

\usepackage[dvipsnames]{xcolor}
\usepackage{hyperref}
\hypersetup{
	colorlinks=true,
	linkcolor=MidnightBlue,
	citecolor=MidnightBlue,
	urlcolor=BrickRed,
}

\usepackage{amsmath}
\allowdisplaybreaks
\usepackage{amssymb}
\usepackage{mathtools}
\usepackage{bbm}
\usepackage{mathrsfs}
\usepackage{euscript}
\usepackage{tikz-cd}
\tikzcdset{ampersand replacement=\&}

\renewcommand{\rm}[1]{\mathrm{#1}}
\newcommand{\id}{\rm{id}}
\newcommand{\e}{\rm{e}}
\renewcommand{\i}{\rm{i}}
\newcommand{\bb}[1]{\mathbbm{#1}}
\newcommand{\eu}[1]{\EuScript{#1}}
\newcommand{\scr}[1]{\mathscr{#1}}
\renewcommand{\cal}[1]{\mathcal{#1}}
\renewcommand{\frak}[1]{\mathfrak{#1}}

\DeclarePairedDelimiterX{\abs}[1]{\lvert}{\rvert}{#1}
\DeclarePairedDelimiterX{\norm}[1]{\lVert}{\rVert}{#1}
\DeclarePairedDelimiterX{\ip}[2]{\langle}{\rangle}{#1,#2}
\DeclarePairedDelimiterX{\ev}[1]{\langle}{\rangle}{#1}
\DeclarePairedDelimiterX{\pair}[2]{\prec}{\succ}{\,#1,#2\,}

\newcommand{\dd}[1]{\,\rm{d}#1}

\newcommand{\ctsp}[2][]{\rm{T}_{#1}^*{#2}}
\newcommand{\loc}{\rm{loc}}
\newcommand{\lif}[1]{L^1_\loc(#1)}
\newcommand{\cont}[3][]{C_{#1}^{#2}(#3)}
\renewcommand{\sf}[1]{\cont{\infty}{#1}}
\newcommand{\sfcs}[1]{\cont[c]{\infty}{#1}}
\newcommand{\tfcsd}[1]{\scr{E}(#1)}
\newcommand{\csdist}[1]{\scr{E}'(#1)}

\newcommand{\tfd}[2][]{\scr{D}_{#1}(#2)}
\newcommand{\dist}[1]{\scr{D}'(#1)}

\newcommand{\lie}[3][]{\rm{#2}^{#1}(#3)}

\newcommand{\grad}{\rm{grad}}
\renewcommand{\div}{\rm{div}}
\newcommand{\vol}[1]{\rm{vol}(#1)}
\newcommand{\supp}[1]{\rm{supp}~#1}
\newcommand{\WF}[1]{\rm{WF}(#1)}
\newcommand{\Ad}[2][]{\rm{Ad}^{#1}_{#2}}
\newcommand{\Endo}[1]{\rm{End}(#1)}

\newcounter{stm}
\counterwithin{stm}{section}
\numberwithin{equation}{section}

\newcommand{\auth}[3]{
	\noindent \textsc{#1} \par
	{\small \textit{#2} \par
		\textit{E-mail:} \href{mailto:#3}{\texttt{#3}}}
}

\newenvironment{stm}[1]
{\refstepcounter{stm} \noindent \textbf{\thestm \hspace{2pt} #1.}}
{}

\newcommand{\esym}[1]{\hspace*{\fill}\mbox{$#1$}}

\newenvironment{prf}
{\noindent\textsc{Proof.}}
{\esym{\square}}

\let\stdthebibliography\thebibliography
\let\stdendthebibliography\endthebibliography
\renewenvironment*{thebibliography}[1]
{\stdthebibliography{GLLV11}}
{\stdendthebibliography}

\usepackage{comment}

\begin{document}
	
	\Large \noindent
	\textbf{Microlocal Analysis of a Deformed Quantum Field Theory}
	\vsp[24pt]
	\normalsize
	\auth{Rishabh Ballal}{Institute for Theoretical Physics, Leipzig University}
	{ballal.mathphys@gmail.com}
	\vsp
	\auth{Albert Much}{Institute for Theoretical Physics, Leipzig University}{much@itp.uni-leipzig.de}
	\vsp
	\auth{Rainer Verch}{Institute for Theoretical Physics, Leipzig University}{rainer.verch@uni-leipzig.de}
	\vsp[24pt]
	\normalsize
	\noindent \textit{Abstract}: A deformation technique, known as the warped convolution, takes quantum fields in Minkowski spacetime to quantum fields in noncommutative Minkowski spacetime.  Since a quantum field is an operator-valued regular distribution and the warped convolution is (weakly) an oscillatory integral of Rieffel-type, we prove that the symbol classes introduced by Hörmander admit extensions which are suited to the warped convolutions of scalar quantum field operators. We further show that, if a particular vector state on the undeformed algebra of field operators fulfills the microlocal spectrum condition, then every vector state on the deformed algebra generated by these warped convolutions fulfills the microlocal spectrum condition.
	\vsp
	\hrule

	\section{Introduction}
    
    Due to Heisenberg's uncertainty principle, the precision of localisation of a system in spacetime is inversely proportional to that of its energy-momentum. If a measurement is then so precise as to initiate gravitational collapse, the information of the system will be obscured. This operational obstacle can be overcome by \cite[\S~1]{Doplicher} uncertainty relations of Planckian order on spacetime itself, and these relations are realised by a noncommutative algebra.
	\par
    Let $X_0, \ldots, X_{d-1}$ denote the unbounded self-adjoint position operators in a Hilbert space $\scr{V}$. Imposing the commutation relations $[X_\mu, X_\nu] = \i\, Q_{\mu \nu}\, \id_{\scr{V}}$, where $Q$ is a real and skew-symmetric matrix, results in a \emph{noncommutative Minkowski spacetime} of $d$-dimensions. Using techniques of Weyl quantisation, the scalar quantum field $\Phi$ in the Fock space $\scr{F}$ should now be represented in the tensor product $\scr{V} \otimes \scr{F}$ \cite[\S~6]{Doplicher}, which we denote formally as $\Phi_\otimes$. However, there exists a deformation $\Phi_Q$ (\cite{Akofor}, \cite{Grosse1}) of $\Phi$ in $\scr{F}$ alone such that the commutation relations of $\Phi_\otimes$ and $\Phi_Q$ coincide. In particular, Grosse and Lechner \cite{Grosse1} constructed a unitary map from $\scr{V} \otimes \scr{F}$ to $\scr{F}$, and with it, proved the unitary equivalence of the mentioned quantum fields.
	\par
	Without recourse to the Fock space, this deformation is implemented in the framework of warped convolutions (\cite{Buchholz1}, \cite{Buchholz2}). Let the scalar quantum field $\Phi$ be a regular distribution that is valued as an unbounded operator in a Hilbert space $\scr{H}$. If $U$ is a strongly continuous unitary representation on $\scr{H}$ of the identity component of the Poincaré group $\bb{R}^d \rtimes \rm{O}(1,d-1)$, with $\bb{R}^d$ as the translation subgroup, then the warped convolution $\eu{W}_Q[\Phi(x)]$ of the distributional kernel $\Phi(x)$ acts as
	\begin{align}
		\eu{W}_Q[\Phi(x)]\, \Psi = (2\pi)^{-d} \int_{\bb{R}^d \times \bb{R}^d} \e^{-\i \eta(\theta,\xi)}\, U(Q\theta)\, \Phi(x)\, U(-Q\theta)\, U(\xi)\, \Psi \dd{(\theta,\xi)}
	\end{align}
	on vectors $\Psi$ that lie in the domain of $\Phi(f) = \int \Phi(x)\, f(x) \dd{x}$ and are smooth under the action $\bb{R}^d \ni a \mapsto U(a)\, \Psi \in \scr{H}$ of translations. The Minkowski metric $\eta$ endows, in a weak sense, the warped convolution with the structure of an oscillatory integral of Rieffel-type \cite[Chap.~1]{Rieffel}. In lieu of Rieffel's analysis, his oscillatory integral
	\begin{align}
		\bb{I}_\eta(\frak{s}) = (2\pi)^{-k} \int_{\bb{R}^k \times \bb{R}^k} \e^{-\i \eta(\theta,\xi)}\, \frak{s}(\theta,\xi) \dd{(\theta,\xi)}
	\end{align}
	will be understood as a continuous linear functional on a subspace $S^m_\rho (\bb{R}_\theta^k \times \bb{R}_\xi^k)$ of Hörmander's \emph{symbol classes} $S^m_\rho (\bb{R}_x^s \times (\bb{R}_\theta^k \times \bb{R}_\xi^k))$ \cite[Chap.~I]{Hoermander2}. This is possible since the oscillatory integral $\bb{I}_\eta$ itself (the part excluding the symbol $\frak{s}$) has no dependence on $x$. A notable consequence \cite[Thm.~3.2]{Lechner} is the consideration of $\rho$ in the range $(-1,1]$ as opposed to Hörmander's range $(0,1]$. However, the dependence of symbols on $x$ needs to be re-introduced for the sake of $\Phi$. As per Hörmander's definition, all symbols belong in $\sf{\bb{R}_x^s \times (\bb{R}_\theta^k \times \bb{R}_\xi^k)}$, and so any $S^m_\rho(\bb{R}_\theta^k \times \bb{R}_\xi^k)$-valued distribution on $\bb{R}_x^s$ which integrates a symbol in $S^m_\rho (\bb{R}_x^s \times (\bb{R}_\theta^k \times \bb{R}_\xi^k))$ against functions in $\sfcs{\bb{R}_x^s}$ has an empty wavefront set. Contrariwise, the correlations of field operators within a `physical' state, which arise in the symbols of the products of warped convolutions, are notorious for their singular character and are merely integrable locally. This poses the first main question to be addressed:
	\begin{center}
		\itshape
		Is there an extension of Hörmander's symbol classes for Rieffel's oscillatory integral, such that the symbols are locally integrable on $\bb{R}_x^s$ and smooth on $\bb{R}_\theta^k \times \bb{R}_\xi^k$?
	\end{center}
	An affirmation is given in Section \ref{sec:osc-int} by introducing the \emph{extended symbol classes} $\scr{X}^{m,\rho}_\loc$. These spaces have the structure of $\lif{\bb{R}_x^s}$-modules (Lemma \ref{lem:module}), and the extended symbols induce $S^m_\rho(\bb{R}_\theta^k \times \bb{R}_\xi^k)$-valued regular distributions on $\bb{R}_x^s$ (Proposition \ref{prop:symbolic-dist}), or \emph{symbolic distributions} in short. Since an oscillatory integral composes with a symbolic distribution to define a distribution on $\bb{R}_x^s$, it is a simple consequence of the continuity of oscillatory integrals that the wavefront set of this composition lies within the wavefront set of the symbolic distribution (Theorem \ref{thm:osc-int-wfs}).
	\par
	After the construction of the scalar quantum field $\Phi$ from the Borchers-Uhlmann algebra in Section \ref{sec:scalar-field}, the above constructions are applied in Section \ref{sec:warped-conv} to show that the warped convolution of a field operator is an oscillatory integral; if $\Psi_1$ and $\Psi_2$ belong to a suitable domain of smoothness, then the inner product
	\begin{align}
		\ip{\Psi_1}{U(Q\theta)\, \Phi(x)\, U(-Q\theta)\, U(\xi)\, \Psi_2} 	\end{align}
	and its smeared version 
	\begin{align}
		\ip{\Psi_1}{U(Q\theta)\, \Phi(f)\, U(-Q\theta)\, U(\xi)\, \Psi_2}
	\end{align}
	define an extended symbol and a symbolic distribution, respectively (Theorem \ref{thm:wc}). In the light of Poincaré covariance for warped convolutions, a deformed algebra can only be generated from the warped convolutions of field operators associated with matrices in the orbit $\Sigma_Q:=\{\Lambda Q \Lambda^T : \Lambda \in \eu{L}_+^\uparrow\}$, under the action of the identity component $\eu{L}_+^\uparrow$ of the Poincaré subgroup $\rm{O}(1,d-1)$. Consequently, it becomes necessary to alter the definition of $n$-point distributions for admissible vector states (Definition \ref{def:warped-states}).
	\par
	The interest in wavefront sets stems from the aforementioned correlations of fields, which depend on the state being considered. If $\omega$ is a quasi-free state on the $*$-algebra of the Klein-Gordon field $\Phi$ in a globally hyperbolic spacetime, then it is characterised as being a \emph{physical} state if the singular part of the function $\omega_2(x,y) = \omega(\Phi(x)\, \Phi(y))$ is identical to that of the Minkowski vacuum. Before Radzikowski, this characterisation expressed $\omega_2$ as a sum of the Hadamard parametrix \cite[\S~3.3]{Kay} and a smooth term, and referred to $\omega$ as a \emph{Hadamard state}. With microlocal analysis, Radzikowski proved \cite{Radzikowski} that a quasi-free state $\omega$ is Hadamard if and only if the (smooth) wavefront set of the \emph{two-point distribution}, induced from $\omega_2$ and denoted identically, satisfies
	\begin{align}
		\WF{\omega_2} = \bigl\{(x_1,\zeta_1; x_2,-\zeta_2) \in \eu{N}^+ \times \eu{N}^- : (x_1,\zeta_1) \sim_1 (x_2,\zeta_2) \bigr\},
	\end{align}
	where $\eu{N}^\pm \subset \ctsp{\bb{R}^4}$ represent the bundles of future($+$)/past($-$)-directed null covectors, and the equivalence relation $(x_1,\zeta_1) \sim_1 (x_2,\zeta_2)$ marks the existence of a null geodesic $\gamma$ from $x_1$ to $x_2$ such that $\zeta_1$ is cotangent to $\gamma$ at $x_1$ and $\zeta_2$ is cotangent to $\gamma$ at $x_2$. This also applies to vector-valued quantum fields (\cite{Sahlmann1}, \cite{Sahlmann2}). In the absence of the Klein-Gordon equation, Brunetti \emph{et al.} \cite{Brunetti1} presented a weaker characterisation, designated the \emph{microlocal spectrum condition} ($\mu$SC; Definition \ref{def:mu-sc}), in which
	\begin{align}
		\WF{\omega_2} \subseteq \bigl\{(x_1,\zeta_1; x_2,-\zeta_2) \in \eu{J}^+ \times \eu{J}^- : (x_1,\zeta_1) \sim_2 (x_2,\zeta_2) \bigr\},
	\end{align}
    where $\eu{J}^\pm \subset \ctsp{\bb{R}^4}$ denote the bundles of future($+$)/past($-$)-directed causal covectors, while the equivalence relation $(x_1,\zeta_1) \sim_2 (x_2,\zeta_2)$ denotes the existence of a piecewise smooth curve $\gamma$ from $x_1$ to $x_2$ such that the parallel transport of $\zeta_1$ along $\gamma$ coincides with $\zeta_2$ at $x_2$. Subsequently, an analytic counterpart of the $\mu$SC was developed using analytic wavefront sets in analytic spacetimes (\cite{Strohmaier2}, \cite{Strohmaier3}). In both, (classes of) globally hyperbolic spacetimes and analytic spacetimes, Gérard and Wrochna utilised techniques of pseudo-differential calculus to tackle questions concerning the existence and construction of corresponding Hadamard states (\cite{Gerard1}, \cite{Gerard2}). Here, we will not consider the analytic situation; grounded in Minkowski spacetime, we implement simple methods of microlocal analysis in addressing the second main question:
	\begin{center}
		\itshape
		When does a vector state on the $*$-algebra generated by the warped convolutions \\
		of scalar quantum field operators fulfill the microlocal spectrum condition?
	\end{center}
	Note that every admissible vector state is defined with a vector from the underlying dense domain of the algebra. As it turns out, the criterion for this fulfillment is quite natural and accumulates in the main physical result of this work (Theorem \ref{thm:mu-sc}):
	\vsp
	\noindent \textbf{Theorem.} \\
	{
		\itshape
		If the state $\omega$ on the algebra $\scr{P}(\bb{R}^4)$ of field operators fulfills the $\mu$SC, then so does every admissible vector state on the algebra $\scr{P}(\bb{R}^4,\Sigma_Q)$ of warped field operators.
	}
	\vsp
	Here, $\omega$ is viewed as a vector state on the algebra $\scr{P}(\bb{R}^4)$ of field operators, but is more importantly recognised as the state on the Borchers-Uhlmann algebra which constructs the Gelfand-Naimark-Segal Hilbert space in which the field operators exist. On the other hand, $\scr{P}(\bb{R}^4,\Sigma_Q)$ is the $*$-algebra of field operators deformed under the warped convolutions associated with the orbit $\Sigma_Q = \{\Lambda Q \Lambda^T : \Lambda \in \eu{L}_+^\uparrow\}$. Clearly, $\Sigma_Q$ presupposes the choice of a reference matrix $Q$, which can lead to different properties of the resulting theory; for example, it can impose locality in wedges (\cite{Grosse1}, \cite{Grosse2}, \cite{Buchholz1}, \cite{Buchholz2}) or enable consistent Wick rotations \cite{Grosse3}. The choice of $Q$ in the latter is to ensure that time remains commutative, and has previously been considered by one of the present authors in analysing quantum Dirac fields in Moyal-Minkowski spacetime \cite{Verch2}. This, in itself, is an example of the joint development with Paschke \cite{Paschke} of (locally covariant) quantum field theory in a Lorentzian analogue of Connes' noncommutative geometry, so as to provide a rigorous understanding of the geometry of quantum spacetime. In contrast, the deformation in commutative Minkowksi spacetime does not reflect any mathematical obstacle to the notion of a spacetime point. An imposition of this sort, on physical grounds, would then necessitate a re-definition of the wavefront set, or possibly of the asymptotic correlation spectrum \cite{Verch1}. Here, we maintain the standard definitions and treat the deformation in its own right.

	\section{Distributions and Wavefront Sets}
	\label{sec:wfs}
	
	So as to keep this article self-contained, we recall some basic definitions and results of distribution theory and wavefront sets.
	\par
	First, consider the following spaces of functions. Let $(K_j)_{j \in \bb{N}}$ be an exhaustion of $\bb{R}^s$ by compact sets, that is, a sequence of compact sets such that $K_j$ is contained in the interior of $K_{j+1}$, and $\bb{R}^s = \bigcup_j K_j$. The space $\sf{\bb{R}^s}$ of smooth complex-valued functions on $\bb{R}^s$ is a Fréchet space $\tfcsd{\bb{R}^s}$ equipped with the family of semi-norms
	\begin{align}
		t_{\alpha,K_j}(\psi) := \sup_{x \in K_j}\, \abs{D^\alpha \psi(x)},
		\qquad \psi \in \tfcsd{\bb{R}^s},
	\end{align}
	lending the topology of uniform convergence in all derivatives on compact sets. The space $\sfcs{\bb{R}^s}$ of compactly supported smooth functions on $\bb{R}^s$ is a subspace $\tfd{\bb{R}^s}$ of $\tfcsd{\bb{R}^s}$, endowed with a locally convex topology as the strict inductive limit \cite[\S~II.6]{Schaefer} of the Fréchet spaces
	\begin{align}
		\tfd[K_j]{\bb{R}^s} := \{\psi \in \tfcsd{\bb{R}^s} : \supp{\psi} \subset K_j\},
	\end{align}
	each of which inherits its topology from $\tfcsd{\bb{R}^s}$ \cite[\S~2.1]{Grubb}.
	\vsp
	\begin{stm}{Definition}
		(\cite[\S~3.3, 3.1]{Grubb}, \cite[Defs.~1, 2]{Strohmaier1}) \\
		The topological dual of $\tfcsd{\bb{R}^s}$ is the space $\csdist{\bb{R}^s}$ of \emph{compactly supported distributions}, and the topological dual of $\tfd{\bb{R}^s}$ is the space $\dist{\bb{R}^s}$ of \emph{distributions}.
		\esym{\diamond}%
	\end{stm}
	\vsp
	The inclusion $\tfd{\bb{R}^s} \subset \tfcsd{\bb{R}^s}$ implies $\csdist{\bb{R}^s} \subset \dist{\bb{R}^s}$. Both spaces, $\csdist{\bb{R}^s}$ and $\dist{\bb{R}^s}$, are locally convex under the weak-$*$ topology \cite[Def.~V.1.1]{Conway}. Often, the characterisation in the next proposition is used as the definition of a distribution.
	\vsp
	\begin{stm}{Proposition}
		\label{prop:distribution}
		\cite[Thm.~2.1.4]{Hoermander1} \\
		\itshape
		A linear functional $u$ on $\tfd{\bb{R}^s}$ is a distribution if and only if, for every $j \in \bb{N}$, there exist constants $k_j \in \bb{N}_0$ and $C_j>0$ such that
		\begin{align}
			\label{eq:distribution}
			\abs{u(\psi)} \leq C_j \sum_{\abs{\alpha} \leq k_j} t_{\alpha,K_j}(\psi),
			\qquad \forall\, \psi \in \tfd[K_j]{\bb{R}^s}.
		\end{align}
	\end{stm}
	\vsp[8pt]
	The evaluation of $u \in \dist{\bb{R}^s}$ at $\psi \in \tfd{\bb{R}^s}$ (or $u \in \csdist{\bb{R}^s}$ at $\psi \in \tfcsd{\bb{R}^s}$) will be denoted by the bilinear form $\pair{u}{\psi} := u(\psi)$. A class of examples which offer great insight into the properties of distributions are the \emph{regular} distributions; every locally integrable function $f$ defines via the injective map $\Lambda : \lif{\bb{R}^s} \to \dist{\bb{R}^s}$ a regular distribution
	\begin{align}
		\pair{\Lambda(f)}{\psi} := \int f(x)\, \psi(x) \dd{x},
		\qquad \forall\, \psi \in \tfd{\bb{R}^s},
	\end{align}
	since $\abs{\pair{\Lambda(f)}{\psi}} \leq \norm{f}_{L^1(K_j)}\, t_{0,K_j}(\psi)$ when $\psi \in \tfd[K_j]{\bb{R}^s}$. Denote the derivative as $D^\alpha := (-\i)^{\abs{\alpha}}\, \partial^\alpha$. The problem of interchanging integrals and derivatives for regular distributions generalises to arbitrary distributions as follows.
	\vsp
	\begin{stm}{Lemma}
		\label{lem:derivative-testf}
		\cite[Thm.~4.1.1]{Friedlander} \\
		\itshape
		Let $u \in \dist{\bb{R}^{s_1}}$ and $\psi \in \tfcsd{\bb{R}^{s_1} \!\times \bb{R}^{s_2}}$. If every $y' \in \bb{R}^{s_2}$ is contained in some open set $U_{y'}$ such that $x \mapsto \psi(x,y)$ is supported in a compact set $K_{y'} \subset \bb{R}^{s_1}$ for all $y \in U_{y'}$, then $\pair{u}{\psi(-,y)} \in \tfd{\bb{R}^{s_2}}$ and $D_y^\alpha\, \pair{u}{\psi(-,y)} = \pair{u}{D_y^\alpha\, \psi(-,y)}$.
	\end{stm}
	\vsp
	This result is instrumental in the construction of tensor products of distributions \cite[Thm.~5.1.1]{Hoermander1}, which,  in turn, is useful for the following Schwartz kernel theorem.
	\vsp
	\begin{stm}{Theorem}
		\label{thm:kernel}
		\cite[Thm.~5.1.2]{Hoermander1} \\
		\itshape
		Every distribution $u \in \dist{\bb{R}^{s_1} \!\times \bb{R}^{s_2}}$ induces a continuous linear map $\eu{K} : \tfd{\bb{R}^{s_2}} \to \dist{\bb{R}^{s_1}}$ under the relation
		\begin{align}
			\label{eq:kernel}
			\pair{\eu{K}\psi}{\varphi} = \pair{u}{\varphi \otimes \psi},
		\end{align}
		and conversely, every such continuous linear map $\eu{K}$ admits a unique distribution $u$ for which relation \eqref{eq:kernel} holds.
	\end{stm}
	\vsp
	Let $u \in \csdist{\bb{R}^s}$. Consistent with regular distributions, the Fourier transform $\eu{F} u$ of $u$ is defined as $\pair{\eu{F} u}{\psi} := \pair{u}{\eu{F} \psi}$, for all $\psi \in \tfcsd{\bb{R}^s}$. When restricted to $\tfd{\bb{R}^s}$, the uniqueness of tensor products \cite[Thm.~5.1.1]{Hoermander1} indicates that $\pair{u}{\eu{F} \psi}$ can be rewritten as \cite[Thm.~7.1.14]{Hoermander1}
	\begin{align}
		\pair{u}{\int \psi(\zeta)\, \e^{\i \eta(-,\zeta)} \dd{\zeta}} 
		= \pair{u \otimes \Lambda(\psi)}{\e^{\i \eta(-,-)}} 
		= \int \pair{u}{\e^{\i \eta(-,\zeta)}}\, \psi(\zeta) \dd{\zeta},
	\end{align}
	for all $\psi \in \tfd{\bb{R}^s}$. In this case, the Fourier transform of $u$ is simply understood as a regular distribution induced by the function $\eu{F}u(\zeta) := \pair{u}{\e^{\i \eta(-,\zeta)}}$. As is the case for compactly supported smooth functions, $u$ is smooth if and only if $\eu{F}u$ is decreasing rapidly (\cite[Thm.~7.3.1]{Hoermander1}, \cite[Thm.~6]{Strohmaier1}). This global characterisation admits a refinement to `microlocally' capture the singular behaviour of distributions.
	\vsp
	\begin{stm}{Definition}
		\cite[Def.~11.1.3]{Friedlander} \\
		Let $u \in \dist{\bb{R}^s}$. A point $(x,\zeta') \in \ctsp{\bb{R}^s} \setminus \cal{Z}$, where $\cal{Z}$ denotes the zero section, is a \emph{regular direction} for $u$ if there exist
		\begin{enumerate}
			\item $\varphi \in \tfd{\bb{R}^s}$ with $\varphi(x) \neq 0$, and
			\item an open conic\footnote{A cone $\Gamma \subseteq \ctsp{\bb{R}^s} \setminus \cal{Z}$ is implicitly defined: if $(x,\zeta) \in \Gamma$, then $(x,t\zeta) \in \Gamma$ for all $t>0$.} neighbourhood $\Gamma$ of $\zeta'$ in which
			\begin{align}
				\label{eq:regular-direction}
				\sup_{\zeta \in \Gamma}\, (1+\norm{\zeta})^N\, \abs{\eu{F}(\varphi u)(\zeta)} < \infty,
				\qquad \forall N \in \bb{N}.
			\end{align}
		\end{enumerate}
		The complement of the set of regular directions determines the closed cone
		\begin{align}
			\WF{u} := \bigl\{(x,\zeta') \in \ctsp{\bb{R}^s} \setminus \cal{Z} : (x,\zeta') \text{~is not a regular direction for~} u \bigr\},
		\end{align}
		known as the \emph{(smooth) wavefront set} of $u$.
		\esym{\diamond}%
	\end{stm}
	\vsp
	The wavefront set is a notable tool of `microlocalisation'; first, $\varphi \in \tfd{\bb{R}^s}$ localises a distribution $u \in \dist{\bb{R}^s}$ in the base space and then the Fourier transform $\eu{F}(\varphi u)(\zeta)$ identifies the non-regular directions in the fibre. Note that the second condition of a regular direction is adjusted if $u$ is valued in a Fréchet space with semi-norms $p_j$; one replaces $\abs{\eu{F}(\varphi u)(\zeta)}$ in \eqref{eq:regular-direction} with $p_j(\eu{F}(\varphi u)(\zeta))$, for all $j \in \bb{N}$.
	\par
	Also, the projection $\pi_2(\WF{u}) := \{\zeta : (x,\zeta) \in \WF{u} \text{~for some~} x \in \bb{R}^s\}$ of the wavefront set onto its covectors will prove beneficial in Section \ref{sec:mu-sc}. If $u$ is a compactly supported distribution, this projection is equivalent to the set of directions in which the Fourier transform $\eu{F}u$ is not rapidly decreasing \cite[Prop.~8.1.3]{Hoermander1}.

	\section{Rieffel's Oscillatory Integrals}
	\label{sec:osc-int}
	
	Let $\eta$ be a non-degenerate bilinear form on $\bb{R}^k$ and let $\chi \in \sfcs{\bb{R}_\theta^k \times \bb{R}_\xi^k}$ be such that \\$\chi(0,0)=1$. Rieffel considered the oscillatory integral \cite[Chap.~1]{Rieffel}
	\begin{align}
		\bb{I}_\eta(\frak{s}) := (2\pi)^{-k}\, \lim_{\epsilon \searrow 0}\, \int_{\bb{R}^k \times \bb{R}^k} \e^{-\i \eta(\theta,\xi)}\, \chi(\epsilon \theta, \epsilon \xi)\, \frak{s}(\theta,\xi) \dd{(\theta,\xi)},
	\end{align}
	with technical assumptions on $\frak{s}$ suited to an isometric action of $\bb{R}^k$. In its stead, $\frak{s}$ will belong to a `symbol class' $S^m_\rho(\bb{R}_\theta^k \times \bb{R}_\xi^k)$ in a style similar to Hörmander's \cite[\S~1.1]{Hoermander2}, but with minor adjustments introduced by Lechner and Waldmann \cite[\S~2]{Lechner}.
	\vsp
	\begin{stm}{Definition} \\
		Let $m \in \bb{R}$ and $-1 < \rho \leq 1$. The vector space of all $\frak{s} \in \sf{\bb{R}_\theta^k \times \bb{R}_\xi^k}$ that fulfill
		\begin{align}
			\abs{D_\theta^\beta D_\xi^\alpha\, \frak{s}(\theta,\xi)} \leq C_{\alpha,\beta}\, (1+\norm{(\theta,\xi)})^{m - \rho(\abs{\alpha} + \abs{\beta})}
		\end{align}
		is a \emph{symbol class} $S^m_\rho(\bb{R}_\theta^k \times \bb{R}_\xi^k)$, and $\frak{s}$ is a \emph{symbol} of order $m$ and type $\rho$.
		\esym{\diamond}%
	\end{stm}
	\vsp
	Contrary to Hörmander's definition \cite[Def.~1.1.1]{Hoermander2}, the symbols are assumed to be polynomially bounded in both, $\theta$ and $\xi$, with the same type $\rho$. This is in anticipation of $\bb{R}_\theta^k \times \bb{R}_\xi^k$ being fibres in the bundle $\bb{R}_x^s \times (\bb{R}_\theta^k \times \bb{R}_\xi^k)$, whose introduction will be delayed so as to accommodate lower regularity in the base space $\bb{R}_x^s$. Another point of contrast is the larger domain of $\rho$; its validity will be shown in Theorem \ref{thm:osc-int}.
	\par
	Henceforth, $S^m_\rho(\bb{R}_\theta^k \times \bb{R}_\xi^k)$ will be abbreviated simply as $S^m_\rho$. Every symbol class $S^m_\rho$ is a Fréchet space with the countable family of semi-norms
	\begin{align}
		p_{\alpha,\beta}(\frak{s}) := \sup_{(\theta,\xi) \in \bb{R}^k \times \bb{R}^k}\, \abs{D_\theta^\beta D_\xi^\alpha\, \frak{s}(\theta,\xi)}\, (1+\norm{(\theta,\xi)})^{-m + \rho(\abs{\alpha} + \abs{\beta})}
	\end{align}
	that determine the topology \cite[Chap.~1]{Grigis}. As evident from the definition, symbol classes become larger if $m$ increases or $\rho$ decreases: that is, if $m \leq m'$ and $\rho \geq \rho'$, then $S^m_\rho \subseteq S^{m'}_{\rho'}$. For constant $\rho$, the vector space $\bigcup_m S^m_\rho$ can therefore be equipped with the strict inductive limit topology of Fréchet spaces.
	\vsp
	\begin{stm}{Examples}
		\label{ex:symbols}
		(\cite[\S~7.1]{Grubb}, \cite[\S~3]{Joshi}) \\
		Functions in $\sf{\bb{R}_\theta^k \times \bb{R}_\xi^k}$ that are positively homogeneous of degree $t$ are in $S^t_1$ since
		\begin{align}
			\abs{\frak{s}(\theta,\xi)}
			= \norm{(\theta,\xi)}^t\, \Bigl\lvert \frak{s} \Bigl(\frac{\theta}{\norm{(\theta,\xi)}}, \frac{\xi}{\norm{(\theta,\xi)}} \Bigr) \Bigr\rvert
			\leq C\, (1+\norm{(\theta,\xi)})^t
		\end{align}
		for $\norm{(\theta,\xi)} > 0$. Functions in $\sfcs{\bb{R}_\theta^k \times \bb{R}_\xi^k}$ are obviously in $\bigcap_m S^m_1$.
		\esym{\diamond}%
	\end{stm}
	\vsp
	\begin{stm}{Lemma}
		\label{lem:symbol-props}
		\cite[Prop.~1.1.6]{Hoermander2} \\
		\itshape
		If $\frak{s} \in S^m_\rho$ and $\frak{t} \in S^{m'}_\rho$ are symbols of the same type $\rho$, then
		\begin{enumerate}
			\item the pointwise product $\frak{st}$ belongs to $S^{m+m'}_\rho$, and
			\item the derivatives $D_\theta^\beta D_\xi^\alpha\, \frak{s}$ belong to $S^{m - \rho(\abs{\alpha} + \abs{\beta})}_\rho$.
		\end{enumerate}
	\end{stm}
	\vsp
	Both properties are easily proven. This product bestows upon $\bigcup_m S^m_\rho$ the structure of a topological $*$-algebra, with $S^0_\rho$ being a unital $*$-subalgebra.
	\vsp
	\begin{stm}{Lemma}
		\label{lem:osc-int-conv}
		\cite[\S~1.2]{Hoermander2} \\
		\itshape
		If $m<-2k$, the integral $\int_{\bb{R}^k \times \bb{R}^k} (1+\norm{(\theta,\xi)})^m \dd{(\theta,\xi)}$ is finite.
	\end{stm}
	\vsp[8pt]
	\begin{prf}
		Let $\norm{-}_E$ denote the Euclidean norm and let $\alpha>0$. In spherical coordinates,
		\begin{align}
			\int_{\bb{R}^k \times \bb{R}^k} (1+\alpha \norm{(\theta,\xi)}_E)^m \dd{(\theta,\xi)} 
			= \Omega_{2k-1}(1) \int_0^\infty (1+\alpha r)^m\, r^{2k-1} \dd{r},
		\end{align}
		where $\Omega_{2k-1}(1)$ is the surface measure of the unit $(2k-1)$-sphere. If $t:=-2k-m>0$,
		\begin{align}\nonumber
			\Omega_{2k-1}(1)\, \alpha^{-2k+1} \int_0^\infty (1+\alpha r)^m (\alpha r)^{2k-1} \dd{r}
			&\leq \Omega_{2k-1}(1)\, \alpha^{-2k+1} \int_0^\infty (1+\alpha r)^{-t-1} \dd{r} \\
			&= \Omega_{2k-1}(1)\, \alpha^{-2k} t^{-1}.
		\end{align}
		By equivalence of all norms on a finite-dimensional vector space \cite[Thm.~III.3.1]{Conway}, for every norm $\norm{-}$ on $\bb{R}_\theta^k \times \bb{R}_\xi^k$, there exists $C>0$ such that $C\, \norm{(\theta,\xi)}_E \leq \norm{(\theta,\xi)}$. Consequently, $(1+\norm{(\theta,\xi)})^m \leq (1+C\norm{(\theta,\xi)}_E)^m$ since $m<0$.
	\end{prf}
	\vsp
	It follows that the oscillatory integral $\bb{I}_\eta$ is absolutely convergent when the symbol $\frak{s}$ has order $m<-2k$ since
	\begin{align}
		(2\pi)^k\, \abs{\bb{I}_\eta(\frak{s})} 
		\leq \lim_{\epsilon \searrow 0} \int \abs{\chi(\epsilon \theta, \epsilon \xi)}\, \abs{\frak{s}(\theta,\xi)} \dd{(\theta,\xi)} 
		\leq p_{0,0}(\frak{s}) \int (1+\norm{(\theta,\xi)})^m \dd{(\theta,\xi)}.
	\end{align}
	From the preceding lemmata, $\bb{I}_\eta$ can be extended to a continuous linear functional on $\bigcup_m S^m_\rho$ by iterating a differential operator which lowers the order of the symbol. As the next theorem will illustrate, the standard construction (\cite[Prop.~1.2.2]{Hoermander2}, \cite[\S~3]{Joshi}) derives a partially different result \cite[Thm.~3.2]{Lechner} owing to the bilinear form $\eta$ being positively homogeneous of degree two.
	\vsp
	\begin{stm}{Theorem}
		\label{thm:osc-int} \\
		\itshape
		Let $\eta$ be a non-degenerate bilinear form on $\bb{R}^k$. If $\chi \in \sfcs{\bb{R}_\theta^k \times \bb{R}_\xi^k}$ fulfills $\chi(0,0)=1$, then the oscillatory integral
		\begin{align}
			\bb{I}_\eta(\frak{s}) = (2\pi)^{-k}\, \lim_{\epsilon \searrow 0}\, \int_{\bb{R}^k \times \bb{R}^k} \e^{-\i \eta(\theta,\xi)}\, \chi(\epsilon \theta, \epsilon \xi)\, \frak{s}(\theta,\xi) \dd{(\theta,\xi)}
		\end{align}
		is a continuous linear functional on $\bigcup_m S^m_\rho$ when $-1 < \rho \leq 1$. That is, if $\frak{s}$ is a symbol of order $m$ and type $\rho$, then
		\begin{align}
			\abs{\bb{I}_\eta(\frak{s})} \leq C_h \sum_{\abs{\alpha} + \abs{\beta} \leq h} p_{\alpha,\beta}(\frak{s})
		\end{align}
		for $h > (\rho+1)^{-1} (m+2k)$.
	\end{stm}
	\vsp[8pt]
	\begin{prf}
		The idea is simple but clouded with technicalities. A differential operator will be constructed such that it leaves the exponential term invariant. It can therefore be introduced within the integral, after which partial integration lowers the order of the symbol by Lemma \ref{lem:symbol-props}. Formally, this is depicted as
		\begin{align}
			\int \e^{-\i \eta(\theta,\xi)}\, \frak{s}(\theta,\xi) \dd{(\theta,\xi)}
			= \int \bigl(M \e^{-\i \eta(\theta,\xi)} \bigr)\, \frak{s}(\theta,\xi) \dd{(\theta,\xi)}
			= \int \e^{-\i \eta(\theta,\xi)}\, \bigl(M^* \frak{s}(\theta,\xi) \bigr) \dd{(\theta,\xi)},
		\end{align}
		where $M$ is a first order differential operator and $M^*$ its formal adjoint under partial integration. Iterating this step eventually makes the symbol integrable by Lemma \ref{lem:osc-int-conv}. However, the technicalities arise since $M$ is not defined at the origin in $\bb{R}_\theta^k \times \bb{R}_\xi^k$. One workaround is to separate the integral over a ball centred at the origin and its complement; the outlined method is only applied to the latter, while the contribution of the former can be made arbitrarily small by reducing the size of the ball.
		\par
		Let $B(\delta)$ denote the open ball of radius $\delta$ centred at the origin in $\bb{R}_\theta^k \times \bb{R}_\xi^k$, and $\overline{B}(\delta)$ its closure. Split the integral as $\bb{I}_\eta(\frak{s}) = \bb{I}_\eta(1_{B(\delta)}\, \frak{s}) + \bb{I}_\eta(1_{B(\delta)^c}\, \frak{s})$. As $\chi$ and $\frak{s}$ are smooth, $\bb{I}(1_{B(\delta)}\, \frak{s})$ converges absolutely:
		\begin{align}
			\label{eq:osc-int-1}
			\abs{\bb{I}_\eta(1_{B(\delta)}\, \frak{s})}
			= \abs{\bb{I}_\eta(1_{\overline{B}(\delta)}\, \frak{s})}
			& \leq (2\pi)^{-k}\, p_{0,0}(\frak{s}) \int_{\overline{B}(\delta)} (1+\norm{(\theta,\xi)})^m \dd{(\theta,\xi)} \notag \\
			& \leq (2\pi)^{-k}\, p_{0,0}(\frak{s})\, \vol{\overline{B}(\delta)} \sup_{(\theta,\xi) \in \overline{B}(\delta)}\, (1+\norm{(\theta,\xi)})^m.
		\end{align}
		\par
		In $B(\delta)^c$, the function $\phi := \norm{\eta(-,\xi)}^2 + \norm{\eta(\theta,-)}^2$ is always non-zero because $\eta$ is non-degenerate. If $\Xi = \phi^{-1} (\eta(-,\xi), \eta(\theta,-)) \in \bb{R}_\theta^k \times \bb{R}_\xi^k$, then the differential operator $M := \i\, \Xi \cdot \grad_{(\theta,\xi)}$ leaves $\e^{-\i \eta(\theta,\xi)}$ invariant: 
		\begin{align}
			M \e^{\i \eta(\theta,\xi)} 
			&= \i\, \phi^{-1} \eta(-,\xi) \cdot \grad_{\theta}(\e^{-\i \eta(\theta,\xi)}) + \i\, \phi^{-1} \eta(\theta,-) \cdot \grad_\xi(\e^{-\i \eta(\theta,\xi)}) \notag \\
			&= \phi^{-1} (\norm{\eta(-,\xi)}^2 + \norm{\eta(\theta,-)}^2)\, \e^{-\i \eta(\theta,\xi)} \notag \\
			&= \e^{-\i \eta(\theta,\xi)}.
		\end{align}
		On replacing $\e^{-\i \eta(\theta,\xi)}$ with $M \e^{-\i \eta(\theta,\xi)}$ in $\bb{I}_\eta(1_{B(\delta)^c}\, \frak{s})$, the Leibniz product rule yields a divergence and a formal adjoint operator $M^*$ as
		\begin{align}
			\label{eq:osc-int-4}
			(M \e^{-\i \eta(\theta,\xi)})\, \chi \frak{s}
			&= \i\, \Xi \cdot \grad_{(\theta,\xi)}(\e^{-\i \eta(\theta,\xi)})\, \chi \frak{s} \notag \\
			&= \i\, \div_{(\theta,\xi)}(\e^{-\i \eta(\theta,\xi)} \chi \frak{s}\, \Xi) - \i\, \e^{-\i \eta(\theta,\xi)} \bigl(\Xi \cdot \grad_{(\theta,\xi)}(\chi \frak{s}) + \div_{(\theta,\xi)}(\Xi)\, \chi \frak{s} \bigr) \notag \\
			&=: \i\, \div_{(\theta,\xi)}(\e^{-\i \eta(\theta,\xi)} \chi \frak{s}\, \Xi) - \i\, \e^{-\i \eta(\theta,\xi)} M^*(\chi \frak{s}).
		\end{align}
		The integral of the first term over $B(\delta)^c$ reduces, by the Gauß theorem and compact support of $\chi$, to an integral over the boundary $\partial B(\delta)$. Proper application of the Gauß theorem pertains to the region enclosed by $B(\delta)^c$ and any concentric ball $B(\delta')$ which contains $\supp{\chi}$. The integral over $\partial B(\delta')$ vanishes and thus the limit $\delta' \to \infty$ exists. If $n$ is the inward unit normal vector field and $\!\dd{\sigma}$ the surface measure on $\partial B(\delta)$, then
		\begin{align}
			&\int_{B(\delta)^c} \div_{(\theta,\xi)}(\e^{-\i \eta(\theta,\xi)} \chi(\epsilon \theta, \epsilon \xi)\, \frak{s}(\theta,\xi)\, \Xi(\theta,\xi)) \dd{(\theta,\xi)} \notag \\
			&\qquad\qquad = \int_{\partial B(\delta)} \e^{-\i \eta(\theta,\xi)} \chi(\epsilon \theta, \epsilon \xi)\, \frak{s}(\theta,\xi)\, \bigl(\Xi(\theta,\xi) \cdot n(\theta,\xi) \bigr) \dd{\sigma}(\theta,\xi)
		\end{align}
		and from the compactness of $\partial B(\delta)$, this integral is absolutely convergent:
		\begin{align}
			\label{eq:osc-int-2}
			&(2\pi)^{-k}\, \lim_{\epsilon \searrow 0} \int_{\partial B(\delta)} \abs{\chi(\epsilon \theta, \epsilon \xi)}\, \abs{\frak{s}(\theta,\xi)}\, \abs{\Xi(\theta,\xi) \cdot n(\theta,\xi)} \dd{(\theta,\xi)} \notag \\
			&\qquad \leq (2\pi)^{-k}\, p_{0,0}(\frak{s})\, \Omega_{2k-1}(\delta) \sup_{(\theta,\xi) \in \partial B(\delta)}\, \abs{\Xi(\theta,\xi) \cdot n(\theta,\xi)}\, (1+\norm{(\theta,\xi)})^m.
		\end{align}
		It remains to analyse the integral of the second term in \eqref{eq:osc-int-4}: $\e^{-\i \eta(\theta,\xi)} M^* (\chi \frak{s})$. First, the function $\chi(\epsilon \theta, \epsilon \xi)$ belongs to $\bigcap_m S^m_1$ for every non-zero $\epsilon$, but is considered as an element of $S^0_1$ since it is the smallest symbol class in which $\lim_{\epsilon \searrow 0} \chi(\epsilon \theta, \epsilon \xi)$ converges. Then, the pointwise product $\chi(\epsilon \theta, \epsilon \xi)\, \frak{s}(\theta,\xi)$ remains in $S^m_\rho$. Second, from Example \ref{ex:symbols} and Lemma \ref{lem:symbol-props}, the symbol classes of the coefficients of $M^* = \Xi \cdot \grad_{(\theta,\xi)} + \div_{(\theta,\xi)}(\Xi)$ are computed: $\Xi_j \in S^{-1}_1$ and $\div_{(\theta,\xi)}(\Xi) \in S^{-2}_1$. Thus, $M^*$ maps $\chi \frak{s} \in S^m_\rho$ to $S^{m-(\rho+1)}_\rho$. After an $h$-fold iteration of the above procedure,
		\begin{align}
			(M^*)^h \bigl(\chi(\epsilon \theta, \epsilon \xi)\, \frak{s}(\theta,\xi) \bigr) 
			= \sum_{\abs{\alpha} + \abs{\beta} \leq h} z_{\alpha,\beta}(\epsilon,\theta,\xi)\, D_\theta^\beta D_\xi^\alpha\, \frak{s}(\theta,\xi)
		\end{align}
		belongs to $S^{m-h(\rho+1)}_\rho$. By Lemma \ref{lem:symbol-props}, $D_\theta^\beta D_\xi^\alpha\, \frak{s}$ is in $S^{m - \rho(\abs{\alpha} + \abs{\beta})}_\rho$ and so $z_{\alpha,\beta}$ must be in $S^{\rho(\abs{\alpha} + \abs{\beta}) - h(\rho+1)}_\rho$. With the respective symbol estimates and $h > (\rho+1)^{-1}(m+2k)$, Lemma \ref{lem:osc-int-conv} is applicable and ensures absolute convergence:
		\begin{align}
			\label{eq:osc-int-3}
			&(2\pi)^{-k}\, \lim_{\epsilon \searrow 0} \int_{B(\delta)^c} \abs{(M^*)^h \bigl(\chi(\epsilon \theta, \epsilon \xi)\, \frak{s}(\theta,\xi) \bigr)} \dd{(\theta,\xi)} \notag \\
			&\quad \leq (2\pi)^{-k}\, \lim_{\epsilon \searrow 0} \int_{B(\delta)^c} \sum_{\abs{\alpha} + \abs{\beta} \leq h} \abs{z_{\alpha,\beta}(\epsilon,\theta,\xi)}\, \abs{D_\theta^\beta D_\xi^\alpha\, \frak{s}(\theta,\xi)} \dd{(\theta,\xi)} \notag \\
			&\quad \leq (2\pi)^{-k}\, \lim_{\epsilon \searrow 0} \sum_{\abs{\alpha} + \abs{\beta} \leq h} C_{\alpha,\beta}\, p_{\alpha,\beta}(\frak{s}) \int_{B(\delta)^c} (1+\norm{(\theta,\xi)})^{m-h(\rho+1)} \dd{(\theta,\xi)} \notag \\
			&\quad \leq C'_h \sum_{\abs{\alpha} + \abs{\beta} \leq h} p_{\alpha,\beta}(\frak{s}).
		\end{align}
		\par
		As a result, $\abs{\bb{I}_\eta(\frak{s})} \leq \abs{\bb{I}_\eta(1_{B(\delta)}\, \frak{s})} + \abs{\bb{I}_\eta(1_{B(\delta)^c}\, \frak{s})}$ is bound from above by the sum of \eqref{eq:osc-int-1}, \eqref{eq:osc-int-2}, and \eqref{eq:osc-int-3}. If $p_{0,0}(\frak{s}) \neq 0$, choose $\delta<1$ so that $\norm{(\theta,\xi)}<1$ implies $(1+\norm{(\theta,\xi)})^m < \max(1,2^m)$ in $\overline{B}(\delta)$. Given $\varepsilon>0$, reduce $\delta$ further until $\vol{\overline{B}(\delta)}$ and $\Omega_{2k-1}(\delta)$ are both less than $(2\pi)^k\, \varepsilon/(2 p_{0,0}(\frak{s}) \max(1,2^m))$. It follows that
		\begin{align}
			\abs{\bb{I}_\eta(\frak{s})} < \varepsilon + C_h \sum_{\abs{\alpha} + \abs{\beta} \leq h}\, p_{\alpha,\beta}(\frak{s})
		\end{align}
		since the sum of \eqref{eq:osc-int-1} and \eqref{eq:osc-int-2} is bound from above by $\varepsilon$.
	\end{prf}
	\vsp
	Of course, the inequality before \eqref{eq:osc-int-3} allows exchanging the limit and the integral. The unused condition $\chi(0,0)=1$ then ensures that the choice of $\chi$ does not affect $\bb{I}_\eta$. However, this use of dominated convergence eliminates $\chi$ from the oscillatory integral and explicitly requires $M^*$ to make the symbol absolutely integrable. When the limit is maintained outside the integral, $\bb{I}_\eta$ can be seen as the limit of convergent integrals.
	\par
	As a consequence of the higher degree of homogeneity of $\eta$, this oscillatory integral accommodates symbols of order $m \in \bb{R}$ and type $-1 <\rho \leq 1$. Rieffel considers \cite[Chap.~1]{Rieffel} a space of functions $\bb{R}_\theta^k \times \bb{R}_\xi^k \to \cal{A}$, where $\cal{A}$ is a Fréchet algebra, and then restricts to the dense subspace that is smooth under an isometric action of $\bb{R}^k$. These maps precisely corresponds to the symbol class $S^0_0$ when $\cal{A} = \bb{C}$. However, Rieffel uses the Laplacian, a second order differential operator, in deriving the criterion $h>k$ for $S^0_0$, thereby proving that $h>(\rho+1)^{-1}(m+2k)$ is not optimal.
	\vsp
	\begin{stm}{Proposition}
		\label{prop:trivial-osc-int}
		(\cite[Prop.~3.9]{Lechner}, \cite[Cor.~1.12]{Rieffel}) \\
		\itshape
		If $\frak{s} \in S^m_\rho$ is independent of $\theta$ or $\xi$, that is, either $\frak{s}(\theta,\xi) = \frak{s}(0,\xi)$ for all $\theta \in \bb{R}^k$ or $\frak{s}(\theta,\xi) = \frak{s}(\theta,0)$ for all $\xi \in \bb{R}^k$, then $\bb{I}_\eta(\frak{s}) = \frak{s}(0,0)$.
	\end{stm}
	\vsp[8pt]
	\begin{prf}
		Without loss of generality, assume that $\frak{s}(\theta,\xi) = \frak{s}(0,\xi)$ for all $\theta \in \bb{R}^k$. Let $\chi$ be of the separable form $\chi_1 \otimes \chi_2$ such that
		\begin{align}
			\bb{I}_\eta (\frak{s}) &= (2\pi)^{-k}\, \lim_{\epsilon \searrow 0}\, \int \e^{-\i \eta(\theta,\xi)}\, \chi_1(\epsilon \theta)\, \chi_2(\epsilon \xi)\, \frak{s}(0,\xi) \dd{(\theta,\xi)} \notag \\
			&= (2\pi)^{-k}\, \lim_{\epsilon \searrow 0}\, \epsilon^{-k} \int \eu{F} \chi_1(\epsilon^{-1} \xi)\, \chi_2(\epsilon \xi)\, \frak{s}(0,\xi) \dd{\xi} \notag \\
			&= (2\pi)^{-k}\, \lim_{\epsilon \searrow 0}\, \int \eu{F} \chi_1(\xi)\, \chi_2(\epsilon^2 \xi)\, \frak{s}(0,\epsilon \xi) \dd{\xi}.
		\end{align}
		Then, dominated convergence and the inverse Fourier transform gives
		\begin{align}
			\bb{I}_\eta (\frak{s}) = \chi_2(0)\, \frak{s}(0,0)\, (2\pi)^{-k} \int \eu{F} \chi_1(\xi) \dd{\xi}
			= \chi_1(0)\, \chi_2(0)\, \frak{s}(0,0),
		\end{align}
		and the result follows from the defining condition $\chi(0,0) = \chi_1(0)\, \chi_2(0) = 1$.
	\end{prf}
	\vsp
	Consider symbols on $\bb{R}_x^s \times (\bb{R}_\theta^k \times \bb{R}_\xi^k)$. According to Hörmander's definition \cite[Def.~1.1.1]{Hoermander2}, every symbol would belong in $\sf{\bb{R}_x^s \times (\bb{R}_\theta^k \times \bb{R}_\xi^k)}$. Since the differential \\operator $M$ constructed on $\e^{-\i \eta(\theta,\xi)}$ in the proof of Theorem \ref{thm:osc-int} will not introduce any derivatives in $x$, the assumption of smoothness in $x$ is unnecessary. Moreover, regular distributions on $\bb{R}_x^s$ induced by these symbols will have empty wavefront sets. Hence, it seems wise to reduce smoothness in $x$ to local integrability. For this, let $(K_j)_{j \in \bb{N}}$ be an exhaustion of $\bb{R}_x^s$ by compact sets.
	\vsp
	\begin{stm}{Definition}
		\label{def:extended-symbols} \\
		Let $m \in \bb{R}$ and $-1 < \rho \leq 1$. The vector space $\lif{\bb{R}_x^s;\, S^m_\rho(\bb{R}_\theta^k \times \bb{R}_\xi^k)}$ is designated an \emph{extended symbol class} $\scr{X}^{m,\rho}_\loc$, with a Fréchet topology determined by the semi-norms
		\begin{align}
			q_{\alpha,\beta,K_j}(\frak{u}) := \int_{K_j} p_{\alpha,\beta}(\frak{u}(x)) \dd{x},
		\end{align}
		where $\frak{u} \in \scr{X}^{m,\rho}_\loc$ is an \emph{extended symbol} of order $m$ and type $\rho$.
		\esym{\diamond}%
	\end{stm}
	\vsp
	If $m \leq m'$ and $\rho \geq \rho'$, then $S^m_\rho \subseteq S^{m'}_{\rho'}$ implies $\scr{X}^{m,\rho}_\loc \subseteq \scr{X}^{m',\rho'}_\loc$. Thus, $\bigcup_m \scr{X}^{m,\rho}_\loc$ can also be equipped with the strict inductive limit topology of Fréchet spaces when $\rho$ is fixed. That an extended symbol class $\scr{X}^{m,\rho}_\loc$ is indeed an extension of a symbol class $S^m_\rho$ follows trivially from the observation that the elements of $S^m_\rho$ are just functions in $\scr{X}^{m,\rho}_\loc$ which are independent of $x$. The algebraic structure of extended symbol classes carries over from symbol classes: for $\frak{u} \in \scr{X}^{m,\rho}_\loc$ and $\frak{v} \in \scr{X}^{m',\rho}_\loc$, the pointwise product $\frak{u}\frak{v}(x) := \frak{u}(x)\, \frak{v}(x)$ defines an element in $\scr{X}^{m+m',\rho}_\loc$ by Lemma \ref{lem:symbol-props}, providing $\bigcup_m \scr{X}^{m,\rho}_\loc$ with the structure of a topological $*$-algebra and $\scr{X}^{0,\rho}_\loc$ that of a unital $*$-subalgebra.
	\vsp
	\begin{stm}{Lemma}
		\label{lem:module} \\
		\itshape
		Each extended symbol class $\scr{X}^{m,\rho}_\loc$ is an $\lif{\bb{R}_x^s}$-module.
	\end{stm}
	\vsp[8pt]
	\begin{prf}
		The canonical embedding $\lif{\bb{R}_x^s} \hookrightarrow \scr{X}^{0,\rho}_\loc$ as functions constant in $(\theta,\xi)$ and the pointwise product $\scr{X}^{m,\rho}_\loc \times \scr{X}^{m',\rho}_\loc \to \scr{X}^{m+m',\rho}_\loc$ in $\bigcup_m \scr{X}^{m,\rho}_\loc$ together define a ring homomorphism $\lif{\bb{R}_x^s} \to \Endo{\scr{X}^{m,\rho}_\loc}$.
	\end{prf}
	\vsp
	\begin{stm}{Proposition}
		\label{prop:symbolic-dist} \\
		\itshape
		Let $\frak{u} \in \scr{X}^{m,\rho}_\loc$. The map $\widetilde{\Lambda} (\frak{u}) : \tfd{\bb{R}_x^s} \to S^m_\rho(\bb{R}_\theta^k \times \bb{R}_\xi^k)$, given by
		\begin{align}
			\pair{\widetilde{\Lambda}(\frak{u})}{\psi} := \int \frak{u}(x)\, \psi(x) \dd{x},
			\qquad \forall\, \psi \in \tfd{\bb{R}_x^s},
		\end{align}
		is an $S^m_\rho$-valued regular distribution on $\bb{R}_x^s$.
	\end{stm}
	\vsp[8pt]
	\begin{prf}
		Let $\psi \in \tfd[K_j]{\bb{R}_x^s}$. By Lemma \ref{lem:module}, $\frak{u}\psi$ belongs to $\scr{X}^{m,\rho}_\loc$ and so
		\begin{align}
			\label{eq:symbolic-dist-1}
			\int_{K_j} p_{\alpha,\beta}(\frak{u}(x)\, \psi(x)) \dd{x}
			= \int_{K_j} p_{\alpha,\beta}(\frak{u}(x))\, \abs{\psi(x)} \dd{x} 
			\leq q_{\alpha,\beta,K_j}(\frak{u})\, t_{0,K_j}(\psi),
		\end{align}
		implying \cite[Thm.~3]{Thomas} that the pointwise product $\frak{u} \psi$ is Pettis integrable; for each $\Sigma$ in the Borel $\sigma$-algebra of $\bb{R}_x^s$, there exists an element $\widetilde{\frak{u} \psi} \in S^m_\rho$ such that $\varphi(\widetilde{\frak{u} \psi}) = \int_\Sigma \varphi \bigl(\frak{u}(x)\, \psi(x) \bigr) \dd{x}$ for all continuous linear functionals $\varphi$ on $S^m_\rho$. Then, the relation
		\begin{align}
			\label{eq:symbolic-dist-2}
			p_{\alpha,\beta}(\pair{\widetilde{\Lambda}(\frak{u})}{\psi}) 
			\leq \int p_{\alpha,\beta} \bigl(\frak{u}(x)\, \psi(x) \bigr) \dd{x}
		\end{align}
		only holds if $p_{\alpha,\beta}(\frak{u}(x)\, \psi(x))$ is measurable, which is true by the definition of extended symbols. Following \eqref{eq:symbolic-dist-2} with \eqref{eq:symbolic-dist-1} hence proves, under Proposition \ref{prop:distribution}, that $\widetilde{\Lambda}(\frak{u})$ is an $S^m_\rho$-valued regular distribution.
	\end{prf}
	\vsp
	\begin{stm}{Definition}
		\label{def:symbolic-dist} \\
		Every $S^m_\rho$-valued regular distribution $\widetilde{\Lambda} (\frak{u})$ on $\bb{R}_x^s$ induced by an extended symbol $\frak{u} \in \scr{X}^{m,\rho}_\loc$ is a \emph{symbolic distribution} of order $m$ and type $\rho$.
		\esym{\diamond}%
	\end{stm}
	\vsp
	Extended symbols and symbolic distributions bestow upon the oscillatory integral, in two ways, the structure of a distribution. First, the function $\bb{I}_\eta \circ \frak{u} : \bb{R}_x^s \to \bb{C}$ is locally integrable and determines a regular distribution $\Lambda(\bb{I}_\eta \circ \frak{u})$ on $\bb{R}_x^s$. Second, the composition $\bb{I}_\eta \circ \widetilde{\Lambda}(\frak{u}) : \tfd{\bb{R}_x^s} \to \bb{C}$ is a distribution on $\bb{R}_x^s$ by Proposition \ref{prop:distribution} since
	\begin{align}
		\abs{\pair{\bb{I}_\eta \circ \widetilde{\Lambda}(\frak{u})}{\psi}} 
		&= \abs{\bb{I}_\eta(\pair{\widetilde{\Lambda}(\frak{u})}{\psi})} \notag \\
		&\leq C_h \sum_{\abs{\alpha} + \abs{\beta} \leq h} p_{\alpha,\beta}(\pair{\widetilde{\Lambda}(\frak{u})}{\psi}) \notag \\
		&\leq C_h\, t_{0,K_j}(\psi) \sum_{\abs{\alpha} + \abs{\beta} \leq h} q_{\alpha,\beta,K_j}(\frak{u}),
		\qquad \forall \psi \in \tfd[K_j]{\bb{R}_x^s},
	\end{align}
	where the first inequality follows from continuity and the second from the inequalities \eqref{eq:symbolic-dist-2} and \eqref{eq:symbolic-dist-1}. As the next proposition will show, both distributions are equal and the oscillatory integral, in a heuristic sense, behaves as an intertwiner of regular and symbolic distributions.
	\vsp
	\begin{stm}{Proposition}
		\label{prop:intertwiner} \\
		\itshape
		If $\frak{u} \in \scr{X}^{m,\rho}_\loc$, then $\bb{I}_\eta \circ \widetilde{\Lambda}(\frak{u}) = \Lambda(\bb{I}_\eta \circ \frak{u})$. That is, the following diagram commutes.
		\begin{equation}
			\begin{tikzcd}[row sep=large]
				\& S^m_\rho(\bb{R}_\theta^k \times \bb{R}_\xi^k) \arrow[dr, "\bb{I}_\eta"] \& \\
				\tfd{\bb{R}_x^s} \arrow[ur,"\widetilde{\Lambda}(\frak{u})"] \arrow[rr,"\Lambda(\bb{I}_\eta \circ\, \frak{u})"] \& \& \bb{C}
			\end{tikzcd}
		\end{equation}
	\end{stm}
	\vsp[8pt]
	\begin{prf}
		The functional $\delta_{(\theta,\xi)}(\frak{s}) := (1+\norm{(\theta,\xi)})^{-m}\, \frak{s}(\theta,\xi)$ on $S^m_\rho$ evaluates $\frak{s}$ at $(\theta,\xi)$ analogous to the $\delta$-distribution and is clearly continuous since $\abs{\delta_{(\theta,\xi)}(\frak{s})} \leq p_{0,0}(\frak{s})$. By the Pettis integrability shown in the proof of Proposition \ref{prop:symbolic-dist},
		\begin{align}
			\pair{\widetilde{\Lambda}(\frak{u})}{\psi}(\theta,\xi)
			&= (1+\norm{(\theta,\xi)})^m\, \delta_{(\theta,\xi)}(\pair{\widetilde{\Lambda}(\frak{u})}{\psi}) \notag \\
			&= \int (1+\norm{(\theta,\xi)})^m\, \delta_{(\theta,\xi)}(\frak{u}(x))\, \psi(x) \dd{x}
			= \int \frak{u}(x)(\theta,\xi)\, \psi(x) \dd{x},
		\end{align}
		for all $\psi \in \tfd{\bb{R}_x^s}$. Applying this relation along with Fubini's theorem gives
		\begin{align}
			\pair{\bb{I}_\eta \circ \widetilde{\Lambda}(\frak{u})}{\psi}
			&= (2\pi)^{-k}\, \lim_{\epsilon \searrow 0} \int \e^{-\i \eta(\theta,\xi)} \chi(\epsilon \theta, \epsilon \xi)\, \pair{\widetilde{\Lambda}(\frak{u})}{\psi}(\theta,\xi) \dd{(\theta,\xi)} \notag \\
			&= (2\pi)^{-k}\, \lim_{\epsilon \searrow 0} \int \e^{-\i \eta(\theta,\xi)} \chi(\epsilon \theta, \epsilon \xi)\, \Bigl(\int \frak{u}(x)(\theta,\xi)\, \psi(x) \dd{x} \Bigr) \dd{(\theta,\xi)} \notag \\
			&= (2\pi)^{-k}\, \lim_{\epsilon \searrow 0} \int \Bigl(\psi(x) \int \e^{-\i \eta(\theta,\xi)} \chi(\epsilon \theta, \epsilon \xi)\, \frak{u}(x)(\theta,\xi) \dd{(\theta,\xi)} \Bigr) \dd{x}.
		\end{align}
		In deriving \eqref{eq:osc-int-3}, the limit $\epsilon \searrow 0$ played no part and so the estimate
		\begin{align}
			(2\pi)^{-k}\, \abs{\psi(x) \int \e^{-\i \eta(\theta,\xi)} \chi(\epsilon \theta, \epsilon \xi)\, \frak{u}(x)(\theta,\xi) \dd{(\theta,\xi)}}
			\leq \abs{\psi(x)}\, C_h \sum_{\abs{\alpha} + \abs{\beta} \leq h} p_{\alpha,\beta}(\frak{u}(x))
		\end{align}
		still holds. Since the upper bound is integrable, dominated convergence results in
		\begin{align}
			\pair{\bb{I}_\eta \circ \widetilde{\Lambda}(\frak{u})}{\psi}
			&= (2\pi)^{-k}\, \lim_{\epsilon \searrow 0} \int \Bigl(\psi(x) \int \e^{-\i \eta(\theta,\xi)} \chi(\epsilon \theta, \epsilon \xi)\, \frak{u}(x)(\theta,\xi) \dd{(\theta,\xi)} \Bigr) \dd{x} \notag \\
			&= \int \Bigl((2\pi)^{-k}\, \lim_{\epsilon \searrow 0} \int \e^{-\i \eta(\theta,\xi)} \chi(\epsilon \theta, \epsilon \xi)\, \frak{u}(x)(\theta,\xi) \dd{(\theta,\xi)} \Bigr)\, \psi(x) \dd{x} \notag \\
			&= \pair{\Lambda(\bb{I}_\eta \circ \frak{u})}{\psi}
		\end{align}
		for all $\psi \in \tfd{\bb{R}_x^s}$. Due to the separating nature of the semi-norms on $\dist{\bb{R}_x^s}$, the desired equivalence is obtained.
	\end{prf}
	\vsp
	The extension of symbol classes to $\bb{R}_x^s \times (\bb{R}_\theta^k \times \bb{R}_\xi^k)$ required local integrability in $x$ and symbolic estimates in $(\theta,\xi)$. In lieu of the introduced extended symbol classes $\lif{\bb{R}_x^s;\, S^m_\rho(\bb{R}_\theta^k \times \bb{R}_\xi^k)}$, these conditions can also be realised with the vector-valued symbols $S^m_\rho(\bb{R}_\theta^k \times \bb{R}_\xi^k) \to \lif{\bb{R}_x^s}$. Lechner and Waldmann \cite{Lechner} analysed symbols that are more generally valued in sequentially complete locally convex spaces, and fed to an oscillatory integral which is then valued in the same space. Here, the extended symbol classes $\scr{X}_\loc^{m,\rho}$ are preferred owing to their interesting algebraic structure and since the oscillatory integral $\bb{I}_\eta$ does not itself require any modification.
	\par
	Finally, the distributional structures introduced with oscillatory integrals permit a little discussion about wavefront sets. To be precise, oscillatory integrals (or any other continuous linear functional on $S^m_\rho$, since the following proof only requires continuity) never enlarge the wavefront set of symbolic distributions.
	\vsp
	\begin{stm}{Theorem}
		\label{thm:osc-int-wfs} \\
		\itshape
		If $\frak{u} \in \scr{X}^{m,\rho}_\loc$, then $\WF{\Lambda(\bb{I}_\eta \circ \frak{u})} = \WF{\bb{I}_\eta \circ \widetilde{\Lambda}(\frak{u})} \subseteq \WF{\widetilde{\Lambda}(\frak{u})}$.
	\end{stm}
	\vsp[8pt]
	\begin{prf}
		The equality follows immediately from Proposition \ref{prop:intertwiner}. For the inclusion, let $(x,\zeta')$ be a regular direction for $\widetilde{\Lambda}(\frak{u})$. Then there exist $\psi \in \tfd{\bb{R}_x^s}$ with $\psi(x) \neq 0$ and an open conic neighbourhood $\Gamma$ of $\zeta'$ in which
		\begin{align}
			\sup_{\zeta \in \Gamma}\, (1+\norm{\zeta})^N p_{\alpha,\beta} (\eu{F}[\psi \widetilde{\Lambda}(\frak{u})](\zeta)) < \infty, 
			\qquad \forall\, N \in \bb{N}.
		\end{align}
		By the continuity of the oscillatory integral (Theorem \ref{thm:osc-int}), there exists $h$ such that
		\begin{align}
			\abs{\eu{F}[\psi(\bb{I}_\eta \circ \widetilde{\Lambda}(\frak{u}))](\zeta)} 
			&= \abs{\bb{I}_\eta(\pair{\widetilde{\Lambda}(\frak{u})}{\e^{\i \eta(-,\zeta)} \psi})} \notag \\
			&\leq C_h \sum_{\abs{\alpha} + \abs{\beta} \leq h} p_{\alpha,\beta}(\pair{\widetilde{\Lambda}(\frak{u})}{\e^{\i \eta(-,\zeta)} \psi}),
		\end{align}
		and the left hand side rapidly decreases in $\Gamma$ since $\pair{\widetilde{\Lambda}(\frak{u})}{\e^{\i \eta(-,\zeta)} \psi} = \eu{F}[\psi \widetilde{\Lambda}(\frak{u})](\zeta)$. Thus, $(x,\zeta')$ is a regular direction for $\bb{I}_\eta \circ \widetilde{\Lambda}(\frak{u})$.
	\end{prf}
	\vsp
	The implications of these results will become apparent in their application to the warped convolution of quantum field operators, which are (weak) oscillatory integrals of Rieffel-type. Explicitly, the extended symbol and symbolic distribution are given in Theorem \ref{thm:wc}, while the wavefront set inclusion offered by the action of the oscillatory integral on the symbolic distribution (in the case of a product of warped convolutions) will initiate the proof of Theorem \ref{thm:mu-sc}.

	\section{The Scalar Quantum Field}
	\label{sec:scalar-field}
	
	Consider $\bb{R}^d$ to be Minkowski spacetime with metric $\eta = \rm{diag}(+1,-1,\ldots,-1)$. Let $\eu{P}$ denote the Poincaré group $\bb{R}^d \rtimes \lie{O}{1,d-1}$ of symmetries for Minkowski spacetime, and $\eu{P}_+^\uparrow$ the component connected to its identity element. Then the \emph{Borchers-Uhlmann algebra} \cite{Borchers, Uhlmann} is defined as
	\begin{align}
		\scr{B}(\bb{R}^d) := \bigoplus_{n=0}^\infty \bigotimes^n \tfd{\bb{R}^d},
	\end{align}
	where $\bigotimes^0 \tfd{\bb{R}^d} := \bb{C}$. An element $f \in \scr{B}(\bb{R}^d)$ is a sequence $(f^{(n)})_{n \in \bb{N}_0}$ with finitely many non-zero components, and the operations on $\scr{B}(\bb{R}^d)$ can easily be extrapolated from linear (or anti-linear, for the involution) and component-wise extension of
	\begin{enumerate}
		\item the associative product $(f \times g)^{(n)} = \sum_{j=0}^n f^{(j)} \otimes g^{(n-j)}$,
		\item the involution $(f_1 \otimes \cdots \otimes f_n)^* = \overline{f_n} \otimes \cdots \otimes \overline{f_1}$ by complex conjugation, and
		\item for every $(a,\Lambda) \in \eu{P}$, the $*$-automorphism
		\begin{align}
			\label{eq:bu-covariance}
			\alpha_{(a,\Lambda)} (f_1 \otimes \cdots \otimes f_n) = (\alpha_{(a,\Lambda)}\, f_1 \otimes \cdots \otimes \alpha_{(a,\Lambda)}\, f_n),
		\end{align}
		where $(\alpha_{(a,\Lambda)}\, f_j) (x) := f_j(\Lambda^{-1}(x-a))$.
	\end{enumerate}
	The product fulfills the involutive property $(f \times g)^* = g^* \times f^*$ and admits an identity element $1_\scr{B} := (1,0,\ldots)$. Hence, $\scr{B}(\bb{R}^d)$ is a unital $*$-algebra \cite[\S~2.1]{Schmuedgen1}. Barring the action of the Poincaré group, the structure of $\scr{B}(\bb{R}^d)$ is that of a tensor algebra \cite[\S~6.1]{Schmuedgen1} over $\tfd{\bb{R}^d}$. On providing $\bigotimes^n \tfd{\bb{R}^d}$ with the projective topology, so as to ensure that the multilinear maps $\otimes_n : (f_1, \ldots, f_n) \mapsto f_1 \otimes \cdots \otimes f_n$ are continuous, $\scr{B}(\bb{R}^d)$ obtains the structure of a topological tensor algebra \cite[\S~6.6]{Schmuedgen1} over $\tfd{\bb{R}^d}$ under the direct sum topology.
	\par
	Let $\omega_n$ be a continuous linear functional on $\bigotimes^n \tfd{\bb{R}^d}$. By the universal property of tensor products, there exists a multilinear map $\overline{\omega}_n : \tfd{\bb{R}^d}^{\times n} \to \bb{C}$ such that $\overline{\omega}_n = \omega_n \circ \otimes_n$. The projective topology on $\bigotimes^n \tfd{\bb{R}^d}$ makes $\overline{\omega}_n$ continuous. When $n=2$, the bilinear map $\overline{\omega}_2$ is clearly equivalent to a linear map $\tfd{\bb{R}^d} \to \dist{\bb{R}^d}$, and the Schwartz kernel theorem (Theorem \ref{thm:kernel}) extends $\overline{\omega}_2$ to a distribution on $\bb{R}^d \times \bb{R}^d$. For $n \geq 2$, a similar iterative argument shows that $\overline{\omega}_n$ is a distribution on $(\bb{R}^d)^{\times n}$. Thus, the functionals $\omega_n$ are known as \emph{$n$-point distributions} and together determine a continuous linear functional $\omega$ on $\scr{B}(\bb{R}^d)$ as
	\begin{align}
		\label{eq:state}
		\omega(f) := \sum_{n=0}^\infty \pair{\omega_n}{f^{(n)}}.
	\end{align}
	This is a \emph{state} on the $*$-algebra $\scr{B}(\bb{R}^d)$ if two constraints are imposed: first, $\omega_0(1) = 1$ ensures normalisation $\omega(1_\scr{B}) = 1$, and second, $\omega(f^* \!\times f) \geq 0$ ensures positivity. Assume further that $\omega$ is Poincaré-invariant: $\alpha_{(a,\Lambda)}^* \omega = \omega$ for all $(a,\Lambda) \in \eu{P}$. For a state on a unital $*$-algebra, the following Gelfand-Naimark-Segal (GNS) construction applies.
	\vsp
	\begin{stm}{Theorem}
		(\cite[Thm.~8.9]{Moretti}, \cite[Thm.~4.38]{Schmuedgen1}) \\
		\itshape
		The pair $(\scr{B}(\bb{R}^d), \omega)$ determines a quadruple $(\scr{H}_\omega, \eu{D}_\omega, \pi_\omega, \Psi_\omega)$, where
		\begin{enumerate}
			\item $\scr{H}_\omega$ is a Hilbert space,
			\item $\eu{D}_\omega$ is a dense subspace of $\scr{H}_\omega$,
			\item $\pi_\omega$ is a $*$-representation of $\scr{B}(\bb{R}^d)$ on $\scr{H}_\omega$ as (possibly unbounded) linear operators with a common invariant domain $\eu{D}_\omega$, and
			\item $\Psi_\omega$ is an algebraically cyclic unit vector: $\pi_\omega(\scr{B}(\bb{R}^d))\, \Psi_\omega = \eu{D}_\omega$.
		\end{enumerate}
		Furthermore, the two tuples are related by the equation
		\begin{align}
			\label{eq:gns}
			\omega(f) = \ip{\Psi_\omega}{\pi_\omega(f)\, \Psi_\omega},
			\qquad \forall\, f \in \scr{B}(\bb{R}^d).
		\end{align}
		The quadruple $(\scr{H}_\omega, \eu{D}_\omega, \pi_\omega, \Psi_\omega)$ is unique up to unitary equivalence.
	\end{stm}
	\vsp
	Let $\cal{L}^+(\eu{D}_\omega)$ denote the $*$-algebra of operators $A$ with domain $\eu{D}_\omega$, each of which admits an adjoint $A^*$ whose domain contains $\eu{D}_\omega$, such that the subspace $\eu{D}_\omega$ remains  invariant under $A$ and $A^+ := A^* \!\restriction_{\eu{D}_\omega}$ \cite[Def.~3.1, Lem.~3.2]{Schmuedgen1}. Moreover, $\cal{L}^+(\eu{D}_\omega)$ is locally convex since every $\Psi \in \eu{D}_\omega$ determines a semi-norm $A \mapsto \norm{A\, \Psi}$.
	\vsp
	\begin{stm}{Definition}
		\label{def:quantum-field} \\
		The \emph{scalar quantum field} $\Phi$ is the operator-valued distribution\footnote{For the quantum field, the conventional notation $\Phi(f)$ is preferred to the dual pairing $\pair{\Phi}{f}$.}
		\begin{align}
			\Phi : \tfd{\bb{R}^d} &\to \cal{L}^+(\eu{D}_\omega) \notag \\
			f &\mapsto \Phi(f) := \pi_\omega \bigl((0,f,0,\ldots) \bigr),
		\end{align}
		and the operator $\Phi(f)$, for any $f \in \tfd{\bb{R}^d}$, is referred to as a \emph{field operator}.
		\esym{\diamond}%
	\end{stm}
	\vsp
	From the definition, it is clear that the scalar quantum field is linear: $\Phi(\lambda f+g) = \lambda \Phi(f) + \Phi(g)$ for all $\lambda \in \bb{C}$, and that the field operators are Hermitian: $\Phi(f)^+ = \Phi(\overline{f})$. Hence, the identity $\id_{\scr{H}_\omega}$ and the field operators $\Phi(f)$, for all $f \in \tfd{\bb{R}^d}$, generate the $*$-algebra $\scr{P}(\bb{R}^d) \cong \pi_\omega(\scr{B}(\bb{R}^d))$. A simple computation using \eqref{eq:state}, \eqref{eq:gns}, the product on $\scr{B}(\bb{R}^d)$, as well as the fact that $\pi_\omega$ is a $*$-representation, yields the relation
	\begin{align}
		\pair{\omega_n}{f_1 \otimes \cdots \otimes f_n}
		= \ip{\Psi_\omega}{\Phi(f_1) \ldots \Phi(f_n)\, \Psi_\omega}.
	\end{align}
	Since the products of field operators evaluated within the GNS vector $\Psi_\omega$ correspond precisely to the $n$-point distributions, the state $\omega$ can be viewed as a \emph{vector state} on $\scr{P}(\bb{R}^d)$. Every $\Psi \in \eu{D}_\omega$ of unit norm defines an \emph{admissible vector state $\omega^\Psi$ on $\scr{P}(\bb{R}^d)$} as $\omega^\Psi(-) := \ip{\Psi}{(-)\, \Psi}$, and the $n$-point distributions for $\omega^\Psi$ take the form
	\begin{align}
		\pair{\omega_n^\Psi}{f_1 \otimes \cdots \otimes f_n}
		:= \ip{\Psi}{\Phi(f_1) \ldots \Phi(f_n)\, \Psi}.
	\end{align}
	Finite convex combinations of vector states also define states on $\scr{P}(\bb{R}^d)$.
	\par
	Due to the Poincaré-invariance of $\omega$, the symmetries are implementable on $\scr{H}_\omega$ as a strongly continuous unitary representation $U$ of $\eu{P}_+^\uparrow$ which fulfills, for all $(a,\Lambda) \in \eu{P}_+^\uparrow$,
	\begin{align}
		\begin{gathered}
			\label{eq:covariance}
			U(a,\Lambda)\, \Psi_\omega = \Psi_\omega,
			\hspace{64pt} U(a,\Lambda)\, \eu{D}_\omega \cong \eu{D}_\omega, \\
			U(a,\Lambda)\, \pi_\omega(f)\, U(a,\Lambda)^{-1} = \pi_\omega(\alpha_{(a,\Lambda)}\, f).
		\end{gathered}
	\end{align}
	This establishes the \emph{covariance} of the field: $U(a,\Lambda)\, \Phi(f)\, U(a,\Lambda)^{-1} = \Phi(\alpha_{(a,\Lambda)}\, f)$. For the translation subgroup $\bb{R}^d$ of $\eu{P}_+^\uparrow$, denote the adjoint action of $U(a) := U(a,\id_{\bb{R}^d})$ by $\Ad[U]{(a)}[-] := U(a)\, (-)\, U(-a)$, noting that $U(a)^{-1} = U(-a)$, and let $f_{(a)}(x) := f(x-a)$. With these abbreviations, translational covariance can be stated as
	\begin{align}
		\label{eq:trans-covariance}
		U(a)\, \Psi_\omega &= \Psi_\omega,
		& U(a)\, \eu{D}_\omega &\cong \eu{D}_\omega,
		& \Ad[U]{(a)} [\Phi(f)] &= \Phi(f_{(a)}).
	\end{align}
	By generalising Stone's theorem \cite[Thm.~6.2]{Schmuedgen2} to $d$-parameter unitary groups, the unitary representation of the translation subgroup takes the form $U(a) = \e^{\i \eta(a,P)}$ for a $d$-tuple $P$ of commuting self-adjoint operators. Usually, it is assumed that the joint spectrum of $P$ is supported in the future light cone, but this \emph{spectrum condition} will be replaced in Section \ref{sec:mu-sc} by a microlocal counterpart. Without the spectrum condition, the state $\omega$ need not be the unique Minkowski vacuum state.
	\par
	As is the case for free field theories, $\Phi$ is considered to be a regular distribution in the weak sense. That is, there are pointlike-localised ``operators'' $\Phi(x)$ such that\footnote{For the regular form of the quantum field, the map $\Lambda : \lif{\bb{R}^d} \to \dist{\bb{R}^d}$ will not be used.}
	\begin{align}
		\ip{\Psi_1}{\Phi(f)\, \Psi_2} 
		= \int \ip{\Psi_1}{\Phi(x)\, \Psi_2}\, f(x) \dd{x},
		\qquad \forall\, \Psi_1, \Psi_2 \in \eu{D}_\omega.
	\end{align}
	Translational covariance for $\Phi(x)$ can be written as $\Ad[U]{(a)}[\Phi(x)] = \Phi(x+a)$. However, $\Phi(x)$ is itself \emph{not} a well-defined operator \cite{Wightman} and should instead be interpreted as a sesquilinear form on $\eu{D}_\omega$. Integrating it against functions in $\tfd{\bb{R}^d}$ localises the field in a manner which allows the expression of \emph{local commutativity}: $[\Phi(f), \Phi(g)] = 0$ when the supports of $f$ and $g$ are spacelike separated. For our purpose, local commutativity of the field is not required.
    \par
    Altogether, the assumptions on the field fall short of the Wightman axioms \cite[\S~3.1]{Streater} due to the exclusion of local commutativity and the spectrum condition.

	\section{Warped Algebras and States}
	\label{sec:warped-conv}
	
	Let $Q \in \Endo{\bb{R}^d}$ define a skew-symmetric bilinear form with the metric: $\eta(\theta,Q\xi) = -\eta(\xi,Q\theta)$, and consider the dense subspace $\eu{D}_\omega^\infty := \{\Psi \in \eu{D}_\omega : a \mapsto U(a)\, \Psi \text{~is smooth}\}$.
	\vsp
	\begin{stm}{Definition}
		\label{def:wc}
		(\cite{Buchholz1}, \cite{Buchholz2}) \\
		The \emph{warped convolution} $\frak{w}_Q [A]$ of an operator $A \in \cal{L}^+(\eu{D}_\omega)$ is the sesquilinear form
		\begin{align}
			\frak{w}_Q [A](\Psi_1,\Psi_2) := (2\pi)^{-d} \int_{\bb{R}^d \times \bb{R}^d} \e^{-\i \eta(\theta,\xi)}\, \ip{\Psi_1}{\Ad[U]{(Q\theta)}[A]\, U(\xi)\, \Psi_2} \dd{(\theta,\xi)},
		\end{align}
		where $\Psi_1$ and $\Psi_2$ belong to the domain $\eu{D}_\omega^\infty$ of $\frak{w}_Q[A]$.
		\esym{\diamond}%
	\end{stm}
	\vsp
	By Proposition \ref{prop:trivial-osc-int}, the warped convolution of $\id_{\scr{H}_\omega}$ reduces to the inner product. For a field operator, the warped convolution is well-defined as an oscillatory integral due to the constructions in Section \ref{sec:osc-int}; this is depicted in the next theorem. Note that the function $\chi \in \sfcs{\bb{R}_\theta^d \times \bb{R}_\xi^d}$ will be notationally suppressed until required.
	\vsp
	\begin{stm}{Theorem}
		\label{thm:wc} \\
		\itshape
		If $\Psi_1,\Psi_2 \in \eu{D}_\omega^\infty$, then
		\begin{align}
			\frak{u}_Q^{\Psi_1,\Psi_2}(x)(\theta,\xi) 
			:= \ip{\Psi_1}{\Ad[U]{(Q\theta)}[\Phi(x)]\, U(\xi)\, \Psi_2}
		\end{align}
		is an extended symbol of order 0 and type 0, which induces the symbolic distribution
		\begin{align}
			\pair{\widetilde{\Lambda}(\frak{u}_Q^{\Psi_1,\Psi_2})}{f} (\theta,\xi) 
			= \ip{\Psi_1}{\Ad[U]{(Q\theta)}[\Phi(f)]\, U(\xi)\, \Psi_2},
			\qquad \forall\, f \in \tfd{\bb{R}^d}.
		\end{align}
	\end{stm}
	\vsp[4pt]
	\begin{prf}
		The derivative of $U(a) = \e^{\i \eta(a,P)}$ is generalised from one-parameter unitary groups \cite[Prop.~6.1]{Schmuedgen2} as $D_a^\beta\, U(a) = U(a)\, \eta(-,P)^\beta$ in the strong operator topology. The Leibniz product rule and the skew-symmetry of $Q$ in the metric $\eta$ then give
		\begin{align}
			&\abs{D_\theta^\beta D_\xi^\alpha\, \frak{u}_Q^{\Psi_1,\Psi_2}(x)(\theta,\xi)} \notag \\
			&\quad = \abs{D_\theta^\beta D_\xi^\alpha\, \ip{U(-Q\theta)\, \Psi_1}{\Phi(x)\, U(-Q\theta)\, U(\xi)\, \Psi_2}} \notag \\
			&\quad \leq \sum_{\gamma \leq \beta} C_{\gamma, \beta}\, \abs{\ip{U(-Q\theta)\, (QP)^{\beta-\gamma}\, \Psi_1}{\Phi(x)\, U(-Q\theta)\, (QP)^\gamma\, U(\xi)\, P^\alpha\, \Psi_2}}.
		\end{align}
		Let $\epsilon > 0$. By the characterisation of suprema, there exists $(\theta_s,\xi_s)$ such that
		\begin{align}
			&\sup_{(\theta,\xi) \in \bb{R}^d \times \bb{R}^d}\, \sum_{\gamma \leq \beta} C_{\gamma, \beta}\, \abs{\ip{U(-Q\theta)\, (QP)^{\beta-\gamma}\, \Psi_1}{\Phi(x)\, U(-Q\theta)\, (QP)^\gamma\, U(\xi)\, P^\alpha\, \Psi_2}} \notag \\
			&\quad < \epsilon + \sum_{\gamma \leq \beta} C_{\gamma, \beta}\, \abs{\ip{U(-Q\theta_s)\, (QP)^{\beta-\gamma}\, \Psi_1}{\Phi(x)\, U(-Q\theta_s)\, (QP)^\gamma\, U(\xi_s)\, P^\alpha\, \Psi_2}}.
		\end{align}
		Since $\Psi_1$ and $\Psi_2$ belong in $\eu{D}_\omega^\infty$, so do the vectors $\widetilde{\Psi}_1^{\beta,\gamma,Q} := U(-Q\theta_s)\, (QP)^{\beta-\gamma}\, \Psi_1$ and $\widetilde{\Psi}_2^{\alpha,\gamma,Q} := U(-Q\theta_s)\, (QP)^\gamma\, U(\xi_s)\, P^\alpha\, \Psi_2$. As a result, the semi-norms of $\scr{X}_\loc^{0,0}$ satisfy
		\begin{align}
			q_{\alpha,\beta,K}(\frak{u}_Q^{\Psi_1,\Psi_2})
			&= \int_K p_{\alpha,\beta} \bigl(\frak{u}_Q^{\Psi_1,\Psi_2}(x) \bigr) \dd{x} \notag \\
			&< \int_K \epsilon \dd{x} + \int_K \sum_{\gamma \leq \beta} C_{\gamma, \beta}\, \abs{\ip{\widetilde{\Psi}_1^{\beta, \gamma, Q}}{\Phi(x)\, \widetilde{\Psi}_2^{\alpha,\gamma,Q}}} \dd{x},
		\end{align}
		which is finite due to the regular form of the scalar quantum field $\Phi$.
		\par
		The extended symbol $\frak{u}_Q^{\Psi_1,\Psi_2} \in \scr{X}_\loc^{0,0}$ induces a symbolic distribution $\widetilde{\Lambda}(\frak{u}_Q^{\Psi_1,\Psi_2})$ via Proposition \ref{prop:symbolic-dist}, which takes the form
		\begin{align}
			\pair{\widetilde{\Lambda}(\frak{u}_Q^{\Psi_1,\Psi_2})}{f}(\theta,\xi)
			&= \int_{\bb{R}^d} \frak{u}_Q^{\Psi_1,\Psi_2}(x)(\theta,\xi)\, f(x) \dd{x} \notag \\
			&= \int_{\bb{R}^d} \ip{\Psi_1}{\Ad[U]{(Q\theta)}[\Phi(x)]\, U(\xi)\, \Psi_2}\, f(x) \dd{x} \notag \\
			&= \ip{\Psi_1}{\Ad[U]{(Q\theta)}[\Phi(f)]\, U(\xi)\, \Psi_2},
		\end{align}
		for all $f \in \tfd{\bb{R}^d}$.
	\end{prf}
	\vsp
	Associated with every sesquilinear form $\frak{w}_Q[\Phi(f)]$ is an operator $\eu{W}_Q[\Phi(f)]$, which will also be referred as the warped convolution of a field operator. Its domain consists of vectors $\Psi_2 \in \eu{D}_\omega^\infty$, each of which admit the existence of a vector $\Psi_3 \in \scr{H}_\omega$ such that $\frak{w}_Q[\Phi(f)](\Psi_1,\Psi_2) = \ip{\Psi_1}{\Psi_3}$ for all $\Psi_1 \in \eu{D}_\omega^\infty$. Since $\eu{D}_\omega^\infty$ is dense in $\scr{H}_\omega$, the vector $\Psi_3$ is uniquely determined. The operator is then defined by setting $\eu{W}_Q[\Phi(f)]\, \Psi_2 = \Psi_3$ \cite[Def.~10.4]{Schmuedgen2}, that is,
	\begin{align}
		\label{eq:form-operator}
		\frak{w}_Q[\Phi(f)](\Psi_1,\Psi_2) = \ip{\Psi_1}{\eu{W}_Q[\Phi(f)]\, \Psi_2},
		\qquad \forall\, \Psi_1 \in \eu{D}_\omega^\infty.
	\end{align}
	In fact, the following proposition shows that the covariance of the field $\Phi$ extends the domain of $\eu{W}_Q[\Phi(f)]$ to $\eu{D}_\omega^\infty$ (while the form domain must be strictly larger than the corresponding operator domain, there is no inconsistency as $\eu{D}_\omega^\infty$ is not the maximal form domain) and permits $\Psi_1$ to be any vector in $\scr{H}_\omega$. This is equivalently written as the following integrability condition.
	\vsp
	\begin{stm}{Theorem}
		\label{thm:weak-integral} \\
		\itshape
		If $\Psi \in \eu{D}_\omega^\infty$, then $\eu{W}_Q[\Phi(f)]\, \Psi$ is weakly integrable.
	\end{stm}
	\vsp[4pt]
	\begin{prf}
		Subject to a first order approximation of $U(a) = \e^{\i \eta(a,P)}$, the covariance of the field \eqref{eq:trans-covariance} reduces to the known commutator $[\eta(-,P)^{e_j}, \Phi(f)]\, \Psi = -\Phi(D^{e_j} f)\, \Psi$, where $\{e_j\}$ is the standard basis of $\bb{N}_0^d$. An arbitrary product of the generators can be shifted across the field operators by iterating this expression:
		\begin{align}
			\label{eq:P-commutator}
			\eta(-,P)^\beta\, \Phi(f)\, \Psi = \sum_{\delta \leq \beta} C_{\delta,\beta}\, \Phi(D^\delta f)\, \eta(-,P)^{\beta-\delta}\, \Psi.
		\end{align}
		Now, the Riesz representation theorem \cite[Thm.~I.3.4]{Conway} identifies continuous linear functionals on $\scr{H}_\omega$ with vectors in $\scr{H}_\omega$ under the inner product, which is why it must be shown first that the oscillatory integral
		\begin{align}
			(2\pi)^{-d}\, \lim_{\epsilon \searrow 0} \int_{\bb{R}^d \times \bb{R}^d} \e^{-\i \eta(\theta,\xi)}\, \chi(\epsilon \theta, \epsilon \xi)\, \ip{\Psi'}{\Ad[U]{(Q\theta)}[\Phi(f)]\, U(\xi)\, \Psi} \dd{(\theta,\xi)}
		\end{align}
		is finite for all $\Psi' \in \scr{H}_\omega$. Applying the Leibniz product rule, the derivative $D_a^\beta\, U(a) = U(a)\, \eta(-,P)^\beta$, and the iterated expression \eqref{eq:P-commutator} altogether gives
		\begin{align}
			&\abs{D_\theta^\beta D_\xi^\alpha\, \ip{\Psi'}{U(Q\theta)\, \Phi(f)\, U(-Q\theta)\, U(\xi)\, \Psi}} \notag \\
			&\quad \leq \sum_{\gamma \leq \beta} C_{\gamma, \beta}\, \abs{\ip{\Psi'}{U(Q\theta)\, (QP)^\gamma\, \Phi(f)\, U(-Q\theta)\, (QP)^{\beta-\gamma}\, U(\xi)\, P^\alpha\, \Psi}} \notag \\
			&\quad \leq \sum_{\delta \leq \gamma \leq \beta} C'_{\delta, \gamma, \beta}\, \abs{\ip{U(-\xi)\, \Psi'}{\Ad[U]{(Q\theta - \xi)}[\Phi(D^{T_Q \delta} f)]\, (QP)^{\gamma-\delta}\, (QP)^{\beta-\gamma}\, P^\alpha\, \Psi}} \notag \\
			&\quad \leq \sum_{\delta \leq \beta} C''_{\delta, \beta}\, \abs{\ip{U(-\xi)\, \Psi'}{\Phi(D^{T_Q \delta} f_{(Q\theta - \xi)})\, (QP)^{\beta-\delta}\, P^\alpha\, \Psi}},
		\end{align}
		where $T_Q \in \Endo{\bb{N}_0^d}$ satisfies $(T_Q)_{\mu \nu} = 1$ if $Q_{\mu \nu} \neq 0$ and $(T_Q)_{\mu \nu} = 0$ if $Q_{\mu \nu} = 0$, in the standard bases of $\bb{N}_0^d$ and $\bb{R}^d$. Using the Cauchy-Schwarz inequality yields
		\begin{align}
			&\abs{\ip{U(-\xi)\, \Psi'}{\Phi(D^{T_Q \delta} f_{(Q\theta-\xi)})\, (QP)^{\beta-\delta}\, P^\alpha\, \Psi}} \notag \\
			&\quad \leq \norm{U(-\xi)\, \Psi'}\, \norm{\Phi(D^{T_Q \delta} f_{(Q\theta-\xi)})\, (QP)^{\beta-\delta}\, P^\alpha\, \Psi} \notag \\
			&\quad \leq \norm{\Psi'}\, C'''_j\ t_{T_Q \delta, K_j}(f_{(Q\theta - \xi)})\, \norm{(QP)^{\beta-\delta}\, P^\alpha\, \Psi},
		\end{align}
		by the definition of $\Phi$ as an operator-valued distribution. Since $\Psi$ belongs to $\eu{D}_\omega^\infty$, so does $(QP)^{\beta-\delta}\, P^\alpha\, \Psi$. Moreover, for each $(\theta,\xi)$, the compact set $K_j$ must contain the support of $f_{(Q\theta-\xi)}$, and can be chosen from a compact exhaustion of $\bb{R}^d$ such that it also includes the support of $f$. Then a translation of $f$ within $K_j$ does not affect the suprema of its derivatives, allowing the replacement of $t_{T_Q \delta,K_j}(f_{(Q\theta-\xi)})$ in the above estimate with $t_{T_Q \delta,K_j}(f)$. On taking the supremum over $(\theta,\xi)$, the semi-norms of $S^0_0$ are thus seen to be finite, and by Theorem \ref{thm:osc-int}, so is the oscillatory integral.
		\par
		For every $\epsilon > 0$, the compact support of $\chi$ ensures that the integral
		\begin{align}
			(2\pi)^{-d} \int_{\bb{R}^d \times \bb{R}^d} \e^{-\i \eta(\theta,\xi)}\, \chi(\epsilon \theta, \epsilon \xi)\, \ip{\Psi'}{\Ad[U]{(Q\theta)}[\Phi(f)]\, U(\xi)\, \Psi} \dd{(\theta,\xi)}
		\end{align}
		converges absolutely for all $\Psi' \in \scr{H}_\omega$, and so there exists $\Psi''_\epsilon \in \scr{H}_\omega'' \cong \scr{H}_\omega$ such that this integral equals $\ip{\Psi'}{\Psi''_\epsilon}$ \cite[Lem.~II.3.1]{Diestel}. In the limit $\epsilon \searrow 0$, the convergence of the oscillatory integral ensures the weak convergence of $\Psi''_\epsilon$, and its weak limit is precisely the weak integral of $\eu{W}_Q[\Phi(f)]\, \Psi$.
	\end{prf}
	\vsp
	Since $\Psi'$ is arbitrary in $\scr{H}_\omega$, the weak integral $\eu{W}_Q[\Phi(f)]\, \Psi$ is uniquely determined. The warped convolution of a field operator, abbreviated as a \emph{warped field operator}, is an operator which acts on $\Psi \in \eu{D}_\omega^\infty$ as
	\begin{align}
		\eu{W}_Q[\Phi(f)]\, \Psi = (2\pi)^{-d} \int_{\bb{R}^d \times \bb{R}^d} \e^{-\i \eta(\theta,\xi)}\, \Ad[U]{(Q\theta)}[\Phi(f)]\, U(\xi)\, \Psi \dd{(\theta,\xi)},
	\end{align}
	in the sense of weak oscillatory integrals. By Proposition \ref{prop:intertwiner}, a warped field operator can be determined from the warped convolution of its pointlike localisation $\Phi(x)$:
	\begin{align}
		\label{eq:wc-pointlike}
		\ip{\Psi'}{\eu{W}_Q[\Phi(f)]\, \Psi}
		&= \pair{\bb{I}_\eta \circ \widetilde{\Lambda}(\frak{u}_Q^{\Psi',\Psi})}{f} \notag \\
		&= \pair{\Lambda(\bb{I}_\eta \circ \frak{u}_Q^{\Psi',\Psi})}{f} 
		= \int \ip{\Psi'}{\eu{W}_Q[\Phi(x)]\, \Psi}\, f(x) \dd{x},
	\end{align}
	for all $\Psi' \in \scr{H}_\omega$ and $\Psi \in \eu{D}_\omega^\infty$. As the first step towards the construction of a $*$-algebra of warped field operators, it must be shown that the product of warped field operators is well-defined on the domain $\eu{D}_\omega^\infty$. This is indirectly verified as follows.
	\vsp
	\begin{stm}{Proposition}
		\label{prop:wc-domain} \\
		\itshape
		Every warped field operator $\eu{W}_Q[\Phi(f)]$, with $f \in \tfd{\bb{R}^d}$, leaves $\eu{D}_\omega^\infty$ invariant.
	\end{stm}
	\vsp[8pt]
	\begin{prf}
		Let $\Psi \in \eu{D}_\omega^\infty$. The vector $\eu{W}_Q[\Phi(f)]\, \Psi$ will belong to $\eu{D}_\omega^\infty$ if it belongs to the domain of the derivative $D_a^\beta\, U(a) = U(a)\, \eta(-,P)^\beta$. Due to the covariance of the field \eqref{eq:trans-covariance}, as well as the iterated expression \eqref{eq:P-commutator}, it holds for all $\Psi' \in \scr{H}_\omega$ that
		\begin{align}
			&\ip{\Psi'}{U(a)\, \eta(-,P)^\beta\, \eu{W}_Q[\Phi(f)]\, \Psi} \notag \\
			&\quad = (2\pi)^{-d} \int \e^{-\i \eta(\theta,\xi)}\, \ip{\Psi'}{U(a)\, \eta(-,P)^\beta\, \Ad[U]{(Q\theta)}[\Phi(f)]\, U(\xi)\, U(a)\, \Psi} \dd{(\theta,\xi)} \notag \\
			&\quad = (2\pi)^{-d} \int \sum_{\delta \leq \beta} C_{\delta,\beta}\, \e^{-\i \eta(\theta,\xi)} \ip{\Psi'}{\Ad[U]{(Q\theta)}[\Phi(D^\delta f_{(a)})]\, U(\xi)\, U(a)\, \eta(-,P)^{\beta-\delta}\, \Psi} \dd{(\theta,\xi)} \notag \\
			&\quad = \sum_{\delta \leq \beta} C_{\delta,\beta}\, \ip{\Psi'}{\eu{W}_Q[\Phi(D^\delta f_{(a)})]\, U(a)\, \eta(-,P)^{\beta-\delta}\, \Psi}.
		\end{align}
		This is a finite sum of warped field operators acting on vectors $U(a)\, \eta(-,P)^{\beta-\delta}\, \Psi$ in $\eu{D}_\omega^\infty$, which is well-defined. Since this relation holds for all $\Psi' \in \scr{H}_\omega$, it represents an equality of weak integrals. Hence, the map $a \mapsto U(a)\, \eu{W}_Q[\Phi(f)]\, \Psi$ is smooth.
	\end{prf}
	\vsp
	Also relevant to the construction of the $*$-algebra of warped field operators are the following standard properties of warped convolutions.
	\vsp
	\begin{stm}{Proposition}
		\label{prop:wc-props}
		\cite[Lem.~2.2, Prop.~2.9, Prop.~2.11]{Buchholz2} \\
		\itshape
		Let $A \in \cal{L}^+(\eu{D}_\omega)$ and $\Psi \in \eu{D}_\omega^\infty$. The warped convolution
		\begin{enumerate}
			\item can be restricted, up to a constant factor, to an integral over $\ker{Q}^\perp \!\times \ker{Q}^\perp$,
			\label{prop:wc-kernel}
			\item can equivalently be written as \label{prop:wc-other-form}
			\begin{align}
				\eu{W}_Q[A]\, \Psi = (2\pi)^{-d} \int \e^{-\i \eta(\theta,\xi)}\, U(\xi)\, \Ad[U]{(Q\theta)}[A]\, \Psi \dd{(\theta,\xi)},
			\end{align}
			\item preserves adjoints: $\eu{W}_Q[A^+] \subseteq \eu{W}_Q[A]^*$,
			\label{prop:wc-adjoint}
			\item is covariant in the sense that, 
			\begin{align}
				U(a,\Lambda)\, \eu{W}_Q[A]\, U(a,\Lambda)^{-1} = \eu{W}_{\Lambda Q \Lambda^T} \bigl[U(a,\Lambda)\, A\, U(a,\Lambda)^{-1} \bigr],
			\end{align}
			where $\Lambda^T$ denotes the transpose of $\Lambda$ under $\eta$. Here, $\Lambda Q \Lambda^T$ is denoted as a matrix product in the standard basis of $\bb{R}^d$ rather than a composition of endomorphisms.%
			\label{prop:wc-covariance}
		\end{enumerate}
	\end{stm}
	\vsp
	If the warped convolution of an operator acts on a translation-invariant vector such as $\Psi_\omega$, then it is a redundant deformation of that operator under Proposition \ref{prop:trivial-osc-int}:
	\begin{align}
		\label{eq:trivial-wc}
		\eu{W}_Q[A]\, \Psi_\omega = A\, \Psi_\omega,
		\qquad \forall\, A \in \cal{L}^+(\eu{D}_\omega).
	\end{align}
	Another case of redundancy is when $Q=0$, as is evident from either Proposition \ref{prop:trivial-osc-int} or Proposition \ref{prop:wc-props}\ref{prop:wc-kernel}. Therefore, $Q$ is also known as a \emph{deformation matrix}, and to be consistent, $Q$ will henceforth refer to its matrix representation in the standard basis. However, the covariance of warped field operators, as depicted in Proposition \ref{prop:wc-props}\ref{prop:wc-covariance}, exposes the fallacy of considering just one deformation matrix $Q$ for the warped field operators. Rather, a reference deformation matrix $Q$ must be chosen, and the warped convolutions associated with its orbit $\Sigma_Q := \{\Lambda Q \Lambda^T : \Lambda \in \eu{L}_+^\uparrow\}$ should be incorporated into the constructed $*$-algebra. Note that the proper orthochronous Lorentz group $\eu{L}_+^\uparrow$ is the identity component of the Poincaré subgroup $\rm{O}(1,d-1)$.
	\vsp
	\begin{stm}{Definition} \\
		Let $Q$ be a deformation matrix. In dimensions $d \geq 3$, a \emph{warped algebra} $\scr{P}(\bb{R}^d, \Sigma_Q)$ is the polynomial algebra generated by $\id_{\scr{H}_\omega}$ and the warped field operators
		\begin{align}
			\bigl\{\eu{W}_{Q'}[\Phi(f)] : f \in \tfd{\bb{R}^d},\, Q' \in \Sigma_Q \bigr\},
		\end{align}
		where $\Sigma_Q = \{\Lambda Q \Lambda^T : \Lambda \in \eu{L}_+^\uparrow\}$ is the orbit of $Q$ under the action of $\eu{L}_+^\uparrow$.
		\esym{\diamond}%
	\end{stm}
	\vsp
	After restricting the adjoints of the warped field operators to the domain $\eu{D}_\omega^\infty$, the warped algebra $\scr{P}(\bb{R}^d, \Sigma_Q)$ can be identified as a topological $*$-subalgebra of $\cal{L}^+(\eu{D}_\omega^\infty)$ due to Proposition \ref{prop:wc-props}\ref{prop:wc-adjoint} and the Hermiticity of the field. The natural next step is to consider states on $\scr{P}(\bb{R}^d, \Sigma_Q)$.
	\vsp
	\begin{stm}{Definition}
		\label{def:warped-states} \\
		Every $\Psi \in \eu{D}_\omega^\infty$ of unit norm defines an \emph{admissible vector state $\omega^\Psi$ on $\scr{P}(\bb{R}^d, \Sigma_Q)$} as $\omega^\Psi(-) := \ip{\Psi}{(-)\, \Psi}$, and a map $\omega_n^\Psi : (\Sigma_Q)^{\times n} \to \dist{\bb{R}^{nd}}$ satisfying
		\begin{align}
			\pair{\omega_n^\Psi (Q_1, \ldots, Q_n)}{f_1 \otimes \cdots \otimes f_n}
			= \ip{\Psi}{\eu{W}_{Q_1}[\Phi(f_1)] \ldots \eu{W}_{Q_n}[\Phi(f_n)]\, \Psi},
		\end{align}
		for all $f_1,\ldots,f_n \in \tfd{\bb{R}^d}$, is a \emph{warped $n$-point distribution} for $\omega^\Psi$.
		\esym{\diamond}%
	\end{stm}
	\vsp
	Technically, referring to $\omega_n^\Psi$ as a distribution is a misnomer, because it is only the image of $\omega_n^\Psi$ which is a distribution. Explicitly, (the image of) $\omega_n^\Psi$ takes the form
	\begin{align}
		\label{eq:warped-n-pt}
		\omega_n^\Psi (Q_1, \ldots, Q_n) = \bb{I}_{\bigoplus^n \eta} \circ \widetilde{\Lambda} (\frak{u}_n^\Psi (Q_1, \ldots, Q_n)),
	\end{align}
	where the symbolic distribution subjected to the oscillatory integral is given by
	\begin{align}
		&\pair{\widetilde{\Lambda} (\frak{u}_n^\Psi (Q_1, \ldots, Q_n))}{f_1 \otimes \cdots \otimes f_n}(\theta_1, \ldots, \theta_n, \xi_1, \ldots, \xi_n) \notag \\
		&\, := \ip{\Psi}{\Ad[U]{(Q_1 \theta_1)}[\Phi(f_1)]\, U(\xi_1) \ldots \Ad[U]{(Q_n \theta_n)}[\Phi(f_n)]\, U(\xi_n)\, \Psi} \notag \\
		&\ = \ip{\Psi}{\Phi(f_{1 (Q_1 \theta_1)}) \ldots \Phi(f_{n (Q_n \theta_n + \sum_{k=1}^{n-1} \xi_k)})\, U(\sum_{i=1}^n \xi_i)\, \Psi},
		\label{eq:warped-sym-dist}
	\end{align}
	due to the covariance of the field \eqref{eq:trans-covariance}. Admissible vector states that are pulled-back along the diagonal map $\Sigma_Q \to (\Sigma_Q)^{\times n}$ can be reformulated as states on an algebra deformed by the Rieffel product \cite[Lem.~2.4]{Buchholz2}.

	\section{Microlocal Spectrum Condition}
	\label{sec:mu-sc}
	
	Henceforth, we set $d=4$ as the Minkowski spacetime dimension. Since an admissible vector state is entirely determined by its $n$-point distributions, wavefront sets offer a characterisation of the singular correlations of field operators within that state. After Radzikowski's \cite{Radzikowski} initial contribution, Brunetti \emph{et al.} \cite{Brunetti1} provided a broader description using elements of graph theory.
	\par
	Let $\scr{G}_n$ denote the set of undirected graphs having $n$ vertices $\{v_i\}$ and finitely many edges $\{e_r\}$. An \emph{immersion} $(x,\gamma,k)$ of a graph $G \in \scr{G}_n$ into $\bb{R}^4$ assigns
	\begin{enumerate}
		\item a point $x(v_i)$ in $\bb{R}^4$ to every vertex $v_i$ in $G$,
		\item a piecewise smooth curve $\gamma_r$ in $\bb{R}^4$ which connects the points $x(v_i)$ and $x(v_j)$, to every edge $e_r$ in $G$ which connects the vertices $v_i$ and $v_j$, and
		\item a constant future-directed covector field $k_r$ on each $\gamma_r$.
	\end{enumerate}
	The immersion \emph{instantiates} a point $(x_1,\zeta_1; \ldots; x_n,\zeta_n) \in (\ctsp{\bb{R}^4})^{\times n} \setminus \cal{Z}^{\times n}$ if
	\begin{align}
		\text{(i)~} x_i &= x(v_i), &
		\text{(ii)~} \zeta_i &= \sum_{\substack{\text{$e_r$ between} \\ \text{$v_i$ and $v_{j>i}$}}} k_r(x_i) - \sum_{\substack{\text{$e_r$ between} \\ \text{$v_i$ and $v_{j<i}$}}} k_r(x_i),
	\end{align}
	for $i = 1,\ldots,n$.
	\vsp
	\begin{stm}{Definition}
		\label{def:mu-sc}
		(\cite{Brunetti1}, \cite{Sanders1}) \\
		Let $\omega$ be a state. If the wavefront sets of its $n$-point distributions $\omega_n$ satisfy
		\begin{align}
			\WF{\omega_n} \subseteq \Gamma_n := \bigl\{ &(x_1,\zeta_1; \ldots; x_n,\zeta_n) \in (\ctsp{\bb{R}^4})^{\times n} \setminus \cal{Z}^{\times n} : \notag \\
			&\quad \text{$\exists\, G \in \scr{G}_n$ and an immersion $(x,\gamma,k)$ of $G$ into $\bb{R}^4$ which} \notag \\
			&\quad \text{instantiates the point $(x_1,\zeta_1; \ldots; x_n,\zeta_n)$} \bigr\},
		\end{align}
		then the state $\omega$ fulfills the \emph{microlocal spectrum condition} ($\mu$SC).
		\esym{\diamond}%
	\end{stm}
	\vsp
	It is not difficult to observe from the definition that $\Gamma_2 = \bigl\{(x,\zeta; y,-\zeta) \in \eu{J}^+ \!\times \eu{J}^- \bigr\}$, where $\eu{J}^\pm \subset \ctsp{\bb{R}^4}$ denote the bundles of future($+$)/past($-$)-directed causal covectors. Sometimes, the piecewise smooth curves $\gamma_r$ of graph immersions are restricted to be light-like (or causal), resulting in the $\mu$SC \emph{with light-like} (or \emph{causal}) \emph{immersions}. Due to the non-locality of the warped convolution, admissible vector states on $\scr{P}(\bb{R}^4, \Sigma_Q)$ are not expected to fulfill these stronger versions of the $\mu$SC. However, there exists a sufficient condition under which these states fulfill the $\mu$SC; this answers the question posed in the introduction.
	\vsp
	\begin{stm}{Lemma}
		\label{lem:mu-sc-incl} \\
		\itshape
		Let $n \in \bb{N}$. The wavefront set of the symbolic distribution \eqref{eq:warped-sym-dist} satisfies
		\begin{align}
			\pi_2(\WF{\widetilde{\Lambda} (\frak{u}_n^\Psi (Q_1,\ldots,Q_n))})
			\subseteq \bigcup_{s,t \in \{0,\ldots,N\}} \bigl\{(\zeta_1,\ldots,\zeta_n) \in \bb{R}^{4n} \setminus \{0\} : \qquad \quad \notag \\
			(0,\ldots,0, \zeta_1,\ldots,\zeta_n, 0,\ldots,0) \in \pi_2(\WF{\omega_{s+n+t}}) \bigr\},
		\end{align}
		for some $N \in \bb{N}_0$ and for arbitrary $Q_1,\ldots,Q_n \in \Sigma_Q$.
	\end{stm}
	\vsp[8pt]
	\begin{prf}
		Let $Q_1,\ldots,Q_n \in \Sigma_Q$. If $\theta = (\theta_1,\ldots,\theta_n) \in \bb{R}^{4n}$, $\xi = (\xi_1,\ldots,\xi_n) \in \bb{R}^{4n}$, and $\zeta = (\zeta_1,\ldots,\zeta_n) \in \bb{R}^{4n}$, then the microlocalisation of $\widetilde{\Lambda} (\frak{u}_n^\Psi (Q_1,\ldots,Q_n))$ is
		\begin{align}
			&p_{\alpha,\beta} \bigl(\eu{F} \bigl[(f_1 \otimes \cdots \otimes f_n)\, \widetilde{\Lambda} (\frak{u}_n^\Psi (Q_1,\ldots,Q_n)) \bigr](\zeta_1,\ldots,\zeta_n) \bigr) \notag \\
			&= p_{\alpha,\beta} \bigl(\pair{\widetilde{\Lambda} (\frak{u}_n^\Psi (Q_1,\ldots,Q_n))}{\e^{\i \eta(-,\zeta_1)} f_1 \otimes \cdots \otimes \e^{\i \eta(-,\zeta_n)} f_n} \bigr) \notag \\
			&= \sup_{(\theta,\xi)}\, \bigl\lvert D_\theta^\beta D_\xi^\alpha\, \ip{\Psi}{\Phi((\e^{\i \eta(-,\zeta_1)} f_1)_{(Q_1 \theta_1)}) \ldots \Phi((\e^{\i \eta(-,\zeta_n)} f_n)_{(Q_n \theta_n + \sum_{i=1}^{n-1} \xi_i)})\, U(\xi_+)\, \Psi} \bigr\rvert,
		\end{align}
		where $\xi_+ := \sum_{i=1}^n \xi_i$. By the linearity of the field, the translations of the exponential functions can be extracted from the inner product; for example, the $j$-th element of this $n$-fold product of field operators becomes
		\begin{align}
			\Phi((\e^{\i \eta(-,\zeta_j)} f_j)_{(Q_j \theta_j + \sum_{i=1}^{j-1} \xi_i)})
			= \e^{\i \eta(\theta_j, Q_j \zeta_j)}\, \e^{-\i \sum_{i=1}^{j-1} \eta(\xi_i, \zeta_j)}\, \Phi(\e^{\i \eta(-,\zeta_j)} f_{j (Q_j \theta_j + \sum_{i=1}^{j-1} \xi_i)}).
		\end{align}
		Noting that $\alpha = (\alpha_1,\ldots,\alpha_n) \in \bb{N}_0^{4n}$ and $\beta = (\beta_1,\ldots,\beta_n) \in \bb{N}_0^{4n}$, the derivatives in the semi-norms $p_{\alpha,\beta}$ act via the Leibniz product rule
		\begin{enumerate}
			\item on the exponential prefactors to introduce monomials in $(\zeta_1,\ldots,\zeta_n)$:
			\begin{gather}
				D_{\theta_j}^{\beta_j - \delta_j}\, \e^{\i \eta(\theta_j, Q_j \zeta_j)}
				= \eta(-,Q_j \zeta_j)^{\beta_j - \delta_j}\, \e^{\i \eta(\theta_j, Q_j \zeta_j)}, \\
				D_{\xi_j}^{\kappa_j}\, \e^{-\i \eta(\xi_j, \sum_{i=j+1}^n \zeta_i)}
				= (-1)^{\abs{\kappa_j}}\, \eta(-, \zeta_{j+1} + \cdots + \zeta_n)^{\kappa_j}\, \e^{-\i \eta(\xi_j, \sum_{i=j+1}^n \zeta_i)},
			\end{gather}
			where the multi-indices $\delta = (\delta_1,\ldots,\delta_n) \in \bb{N}_0^{4n}$ and $\kappa = (\kappa_1,\ldots,\kappa_{n-1},0) \in \bb{N}_0^{4n}$ satisfy the constraints $\delta \leq \beta$ and $\kappa \leq \alpha$ respectively,
			\item on the translations of the functions $f_1,\ldots,f_n$ by application of Lemma \ref{lem:derivative-testf}:
			\begin{align}
				\Phi(\e^{\i \eta(-,\zeta_j)}\, D_{\theta_j}^{\delta_j} D_{\xi_1}^{\gamma_{1,j-1}} \ldots D_{\xi_{j-1}}^{\gamma_{j-1,1}}\, f_{j (Q_j \theta_j + \sum_{i=1}^{j-1} \xi_i)}),
			\end{align}
			where $\gamma = \{(\gamma_{i,1}, \ldots, \gamma_{i,n-i}) \in \bb{N}_0^{4(n-i)} : i = 1,\ldots,n-1\}$ satisfies the constraint $\kappa_i + \sum_{l=1}^{n-i} \gamma_{i,l} \leq \alpha_i$, and
			\item on the unitary operator $U(\xi_+) = U(\xi_1) \ldots U(\xi_n)$ in the strong operator topology:
			\begin{align}
				D_{\xi_j}^{\alpha_j - \kappa_j - \sum_{i=1}^{n-j} \gamma_{j,i}}\, U(\xi_j)\, \Psi
				= U(\xi_j)\, \eta(-,P)^{\alpha_j - \kappa_j - \sum_{i=1}^{n-j} \gamma_{j,i}}\, \Psi,
			\end{align}
			since $\Psi \in \eu{D}_\omega^\infty$.
		\end{enumerate}
		After differentiation, the exponential prefactors and the effect of the metric $\eta$ are both redundant when considering the absolute value in the semi-norms $p_{\alpha,\beta}$. Altogether,
		\begin{align}
			\label{eq:mu-sc-ineq}
			&p_{\alpha,\beta} \bigl(\eu{F} \bigl[(f_1 \otimes \cdots \otimes f_n)\, \widetilde{\Lambda} (\frak{u}_n^\Psi (Q_1,\ldots,Q_n)) \bigr](\zeta_1,\ldots,\zeta_n) \bigr) \notag \\
			&\leq \sup_{(\theta,\xi)}\, \sum_{\kappa, \gamma, \delta} C_{\kappa, \gamma, \delta, \alpha, \beta}\, \abs{(Q_1 \zeta_1)^{\beta_1 - \delta_1} \ldots (Q_n \zeta_n)^{\beta_n - \delta_n}\, (\zeta_2 + \cdots + \zeta_n)^{\kappa_1} \ldots (\zeta_n)^{\kappa_{n-1}}} \notag \\
			&\quad\, \lvert \langle \Psi, \Phi(\e^{\i \eta(-,\zeta_1)} D_{\theta_1}^{\delta_1} f_{1 (Q_1 \theta_1)}) \ldots \Phi(\e^{\i \eta(-,\zeta_n)}\, D_{\theta_n}^{\delta_n} D_{\xi_1}^{\gamma_{1,n-1}} \ldots D_{\xi_{n-1}}^{\gamma_{n-1,1}}\, f_{n (Q_n \theta_n + \sum_{i=1}^{n-1} \xi_i)}) \notag \\
			&\qquad \qquad U(\xi_+)\, P^{\alpha_1 - \kappa_1 - \sum_{i=1}^{n-1} \gamma_{1,i}} \ldots P^{\alpha_n}\, \Psi \rangle \rvert.
		\end{align}
		The monomials in $\zeta$ are of finite degree and will not affect the rapid decrease. Thus, the focus is on the inner product in each summand.
		\par
		Since $\Psi_\omega$ is cyclic for $\scr{P}(\bb{R}^4)$, there exist constants $C' \in \bb{C}$ and $N' \in \bb{N}$, as well as functions $h^{(s,1)}, \ldots, h^{(s,s)} \in \tfd{\bb{R}^4}$ for $s = 1,\ldots,N'$ such that
		\begin{align}
			\Psi = C'\, \Psi_\omega + \sum_{s=1}^{N'} \Phi(h^{(s,s)}) \ldots \Phi(h^{(s,1)})\, \Psi_\omega.
		\end{align}
		Similarly, each $\sigma := (\kappa, \gamma, \alpha)$ admits the existence of constants $C''_\sigma \in \bb{C}$ and $N''_\sigma \in \bb{N}$, as well as functions $g_\sigma^{(t,1)}, \ldots, g_\sigma^{(t,t)} \in \tfd{\bb{R}^4}$ for $t = 1,\ldots,N''_\sigma$ such that
		\begin{align}
			&U(\xi_+)\, P^{\alpha_1 - \kappa_1 - \sum_{i=1}^{n-1} \gamma_{1,i}} \ldots P^{\alpha_n}\, \Psi \notag \\
			&\quad = C''_\sigma\, U(\xi_+)\, \Psi_\omega + \sum_{t=1}^{N''_\sigma} U(\xi_+)\, \Phi(g_\sigma^{(t,1)}) \ldots \Phi(g_\sigma^{(t,t)})\, \Psi_\omega \notag \\
			&\quad = C''_\sigma\, \Psi_\omega + \sum_{t=1}^{N''_\sigma} \Phi((g_\sigma^{(t,1)})_{(\xi_+)}) \ldots \Phi((g_\sigma^{(t,t)})_{(\xi_+)})\, \Psi_\omega,
		\end{align}
		due to the covariance of the field and the invariance of $\Psi_\omega$ under translations \eqref{eq:trans-covariance}. Recalling $\bigotimes^0 \tfd{\bb{R}^{4}} := \bb{C}$ from the Borchers-Uhlmann algebra allows the expression of each inner product in \eqref{eq:mu-sc-ineq} as the sum
		\begin{align}
			\label{eq:snt-pt}
			&\sum_{s=0}^{N'} \sum_{t=0}^{N''_\sigma} \prec\! \omega_{s+n+t}, h^{(s)*} \otimes \e^{\i \eta(-,\zeta_1)}\, D_{\theta_1}^{\delta_1} f_{1 (Q_1 \theta'_1)} \otimes \cdots \notag \\
			&\qquad \qquad \quad \otimes \e^{\i \eta(-,\zeta_n)}\, D_{\theta_n}^{\delta_n} D_{\xi_1}^{\gamma_{1,n-1}} \ldots D_{\xi_{n-1}}^{\gamma_{n-1,1}} f_{n (Q_n \theta_n + \sum_{i=1}^{n-1} \xi_i)} \otimes (g_\sigma^{(t)})_{(\xi_+)} \!\succ,
		\end{align}
		where $h^{(s)*} := \overline{h^{(s,1)}} \otimes \cdots \otimes \overline{h^{(s,s)}}$ with $h^{(0)} := C'$, and $(g_\sigma^{(t)})_{(\xi_+)} := (g_\sigma^{(t,1)})_{(\xi_+)} \otimes \cdots \otimes (g_\sigma^{(t,t)})_{(\xi_+)}$ with $(g_\sigma^{(0)})_{(\xi_+)} := C''_\sigma$. In every summand of \eqref{eq:snt-pt}, the Fourier transform is evaluated only with respect to the slots of the functions $f_1,\ldots,f_n$. Hence, it will be useful to restructure the $(s+n+t)$-point distributions.
		\par
		The final part is, in essence, similar to Hörmander's inclusion \cite[Thm.~8.2.12]{Hoermander1} for the wavefront set of a distribution in the context of the Schwartz kernel theorem (Theorem \ref{thm:kernel}). Consider $s,t \in \bb{N}_0$ and $(0,\ldots,0, \zeta'_1,\ldots,\zeta'_n, 0,\ldots,0) \notin \pi_2(\WF{\omega_{s+n+t}})$. By definition, there exists an open conic neighbourhood $\Gamma_{s,t}$ of this point in which
		\begin{align}
			\label{eq:snt-ft}
			&\eu{F} \bigl[(\widetilde{h}^{(s)} \otimes \widetilde{f}_1 \otimes \cdots \otimes \widetilde{f}_n \otimes \widetilde{g}^{(t)})\, \omega_{s+n+t} \bigr](0,\ldots,0, \zeta_1,\ldots,\zeta_n, 0,\ldots,0) \notag \\
			&\qquad = \pair{\omega_{s+n+t}}{\widetilde{h}^{(s)} \otimes \e^{\i \eta(-,\zeta_1)} \widetilde{f}_1 \otimes \cdots \otimes \e^{\i \eta(-,\zeta_n)} \widetilde{f}_n \otimes \widetilde{g}^{(s)}}
		\end{align}
		decreases rapidly, for all $\widetilde{h}^{(s)} \in \bigotimes^s \tfd{\bb{R}^4}$, $\widetilde{f}_1, \ldots, \widetilde{f}_n \in \tfd{\bb{R}^4}$, and $\widetilde{g}^{(t)} \in \bigotimes^t \tfd{\bb{R}^4}$. This can be extended to identify when \eqref{eq:snt-pt} is rapidly decreasing, by accounting for the different values of $s$ and $t$. Set $N$ as the maximum value of the set $\{N'\} \cup \{N''_\sigma : \sigma = (\kappa, \gamma, \alpha)\}$, which must be finite under the implemented cyclicity, and consider a point $(\zeta'_1,\ldots,\zeta'_n)$ from the set
		\begin{align}
			\label{eq:st-set}
			\bigcap_{s,t \in \{0,\ldots,N\}} \bigl\{(\zeta_1,\ldots,\zeta_n) \in \bb{R}^{4n} \setminus \{0\} : (0,\ldots,0, \zeta_1,\ldots,\zeta_n, 0,\ldots,0) \notin \pi_2(\WF{\omega_{s+n+t}}) \bigr\}.
		\end{align}
		Since a finite intersection of open sets is open, there exists\footnote{Under the canonical embeddings of $\Gamma_{s,t} \subset \bb{R}^{4(s+n+t)}$ into $\bb{R}^{4(N+n+N)}$.} an open conic neighbourhood $\Gamma := \bigcap_{s,t \in \{0,\ldots,N\}} \Gamma_{s,t}$ of the point $(\zeta'_1,\ldots,\zeta'_n)$ in which \eqref{eq:snt-pt} decreases rapidly. The dependence of \eqref{eq:snt-pt} on $(\theta,\xi)$ is relevant only to the localisation of $\omega_{s+n+t}$, and has been bypassed by considering, in \eqref{eq:st-set}, only the covectors which never belong to (the projection of) the wavefront set, such that the Fourier transform of $\omega_{s+n+t}$ will rapidly decrease regardless of where it is localised. Consequently, the right hand side of \eqref{eq:mu-sc-ineq} is rapidly decreasing in $\Gamma$, and obviously so is the left hand side, irrespective of $\alpha$ and $\beta$. This implies that $(\zeta'_1,\ldots,\zeta'_n) \notin \pi_2(\WF{\widetilde{\Lambda} (\frak{u}_n^\Psi (Q_1,\ldots,Q_n))})$.
	\end{prf}
	\vsp
	\begin{stm}{Lemma}
		\label{lem:mu-sc-folium}
		\cite[Thm.~4.5]{Brunetti1} \\
		\itshape
		Let $n \in \bb{N}$ and let $s,t \in \bb{N}_0$. If $\omega$ fulfills the $\mu$SC and
		\begin{align}
			(c_1,0; \ldots; c_s,0; a_1,\zeta_1; \ldots; a_n,\zeta_n; b_1,0; \ldots; b_t,0) \in \WF{\omega_{s+n+t}},
		\end{align}
		then there exists a graph $H \in \scr{G}_n$ and an immersion of $H$ into $\bb{R}^4$ which instantiates the point $(a_1,\zeta_1; \ldots; a_n,\zeta_n) \in (\ctsp{\bb{R}^4})^{\times n} \setminus \cal{Z}^{\times n}$.
	\end{stm}
	\vsp[8pt]
	\begin{prf}
		Assume that $s$ and $t$ are both non-zero; the other cases can be proven by an obvious reduction of the following procedure.
		\par
		Since $\omega$ fulfills the $\mu$SC, there exists a graph $G \in \scr{G}_{s+n+t}$ and an immersion $(x,\gamma,k)$ of $G$ into $\bb{R}^4$ which instantiates the point
		\begin{align}
			(c_1,0; \ldots; c_s,0; a_1,\zeta_1; \ldots; a_n,\zeta_n; b_1,0; \ldots; b_t,0) \in (\ctsp{\bb{R}^4})^{\times (s+n+t)} \setminus \cal{Z}^{\times (s+n+t)}.
		\end{align}
		In particular, the following covector relations are satisfied:
		\begin{align}
			0 &= \sum_{\substack{\text{$e_r$ between} \\ \text{$v_i$ and $v_{j>i}$}}} k_r(c_i) - \sum_{\substack{\text{$e_r$ between} \\ \text{$v_i$ and $v_{j<i}$}}} k_r(c_i), &
			i &= 1,\ldots,s, \\
			0 &= \sum_{\substack{\text{$e_r$ between} \\ \text{$v_{s+n+i}$ and} \\ v_{j>s+n+i}}} k_r(b_i) - \sum_{\substack{\text{$e_r$ between} \\ \text{$v_{s+n+i}$ and} \\ v_{j<s+n+i}}} k_r(b_i), &
			i &= 1,\ldots,t.
		\end{align}
		At $c_1$, the covector $0$ is a sum of only future-directed covectors $k_r(c_1)$, each of which must be zero. Due to their constancy, the covector fields $k_r$ vanish along all piecewise smooth curves $\gamma_r$ which start or end at $c_1$. As a result, the covector $0$ at $c_2$ is again a sum of only future-directed covectors $k_r(c_2)$, each of which must be zero. This process can be iterated up to $c_s$, nullifying the covector fields $k_r$ along curves $\gamma_r$ which either start or end at $c_i$, for all $i \in \{1,\ldots,s\}$. Since these covector fields, curves, and points are redundant in the immersion, they can be dropped. Similarly, the covector $0$ at $b_t$ is a sum of only past-directed covectors $k_r(b_t)$, each of which must be zero. Traced in the reverse direction, an identical argument eliminates the covector fields $k_r$ along all curves $\gamma_r$ which either start or end at $b_i$, for all $i \in \{1,\ldots,t\}$, allowing these covector fields, curves, and points to be removed from the immersion. Hence, the graph $G$ can be restricted to the subgraph $H \in \scr{G}_n$, induced by the vertex subset that maps to the point $(a_1,\ldots,a_n)$ under the immersion $(x,\gamma,k)$. In other words, this is an immersion of $H$ which instantiates the point $(a_1,\zeta_1; \ldots; a_n,\zeta_n)$.
	\end{prf}
	\vsp
	\begin{stm}{Theorem}
		\label{thm:mu-sc} \\
		\itshape
		If the state $\omega$ on the algebra $\scr{P}(\bb{R}^4)$ of field operators fulfills the $\mu$SC, then so does every admissible vector state on the algebra $\scr{P}(\bb{R}^4,\Sigma_Q)$ of warped field operators.
	\end{stm}
	\vsp[8pt]
	\begin{prf}
		Let $\omega^\Psi$ be an admissible vector state, and let $Q_1, \ldots, Q_n \in \Sigma_Q$ be arbitrary. From the inclusion offered by Theorem \ref{thm:osc-int-wfs} regarding the wavefront sets of oscillatory integrals acting on symbolic distributions, \eqref{eq:warped-n-pt} gives
		\begin{align}
			\label{eq:mu-sc-incl-1}
			\WF{\omega_n^\Psi (Q_1, \ldots, Q_n)}
			&= \WF{\bb{I}_{\bigoplus^n \eta} \circ \widetilde{\Lambda} (\frak{u}_n^\Psi (Q_1, \ldots, Q_n))} \notag \\
			&\subseteq \WF{\widetilde{\Lambda} (\frak{u}_n^\Psi (Q_1, \ldots, Q_n))} \notag \\
			&\subseteq \bigl\{(x_1,\zeta_1; \ldots; x_n,\zeta_n) \in (\ctsp{\bb{R}^4})^{\times n} \setminus \cal{Z}^{\times n} : \notag \\
			&\qquad (\zeta_1,\ldots,\zeta_n) \in \pi_2(\WF{\widetilde{\Lambda} (\frak{u}_n^\Psi (Q_1, \ldots, Q_n))}) \bigr\},
		\end{align}
		where the last inclusion is a simple consequence of the definition of the wavefront set. Lemma \ref{lem:mu-sc-incl} then provides, for some $N \in \bb{N}_0$, the inclusion
		\begin{align}
			\label{eq:mu-sc-incl-2}
			&\bigl\{(x_1,\zeta_1; \ldots; x_n,\zeta_n) \in (\ctsp{\bb{R}^4})^{\times n} \setminus \cal{Z}^{\times n} : (\zeta_1,\ldots,\zeta_n) \in \pi_2(\WF{\widetilde{\Lambda} (\frak{u}_n^\Psi (Q_1, \ldots, Q_n))}) \bigr\} \notag \\
			&\qquad \qquad \subseteq \bigcup_{s,t \in \{0,\ldots,N\}} \bigl\{(x_1,\zeta_1; \ldots; x_n,\zeta_n) \in (\ctsp{\bb{R}^4})^{\times n} \setminus \cal{Z}^{\times n} : \notag \\
			&\qquad \qquad \qquad \qquad \qquad (0,\ldots,0, \zeta_1,\ldots,\zeta_n, 0,\ldots,0) \in \pi_2(\WF{\omega_{s+n+t}}) \bigr\}.
		\end{align}
		Consider $(x_1,\zeta_1; \ldots; x_n,\zeta_n) \in (\ctsp{\bb{R}^4})^{\times n} \setminus \cal{Z}^{\times n}$ such that $(0,\ldots,0, \zeta_1,\ldots,\zeta_n, 0,\ldots,0)$ belongs to $\pi_2(\WF{\omega_{s+n+t}})$ for some $s,t \in \{0,\ldots,N\}$. This implies the existence of a point $(c_1,\ldots,c_s, a_1,\ldots,a_n, b_1,\ldots,b_t) \in \bb{R}^{4(s+n+t)}$ such that
		\begin{align}
			(c_1,0; \ldots; c_s,0; a_1,\zeta_1; \ldots; a_n,\zeta_n; b_1,0; \ldots; b_t,0) \in \WF{\omega_{s+n+t}},
		\end{align}
		and, by Lemma \ref{lem:mu-sc-folium}, there exist a graph $H \in \scr{G}_n$ and an immersion of $H$ into $\bb{R}^4$ which instantiates the point $(a_1,\zeta_1; \ldots; a_n,\zeta_n)$.
		\par
		Let $\overline{H} \in \scr{G}_n$ be a reduced version of $H$; it has the same vertex set, but admits only one edge between every pair of vertices connected by one or more edges in $H$. Define an immersion $(x',\gamma',k')$ of $\overline{H} = (\{v_i\}, \{\overline{e}_r\})$ into $\bb{R}^4$ in terms of $H = (\{v_i\}, \{e_r\})$ as
		\begin{enumerate}
			\item $x'(v_i) = x_i$,
			\item $\gamma'_r$ is the straight line connecting $x'(v_i)$ and $x'(v_j)$ if $\overline{e}_r$ connects $v_i$ and $v_j$,
			\item $k'_r$ is the parallel transport along $\gamma'_r$ of $k'_r(x'(v_i)) = \sum_{\text{$e_r$ between $v_i$ and $v_j$}} k_r(x(v_i))$, where $v_i$ and $v_j$ are the vertices corresponding to the endpoints of $\gamma'_r$.
		\end{enumerate}
		As this immersion is constructed precisely to instantiate the point $(x_1,\zeta_1; \ldots; x_n,\zeta_n)$, it can be concluded that
		\begin{align}
			\label{eq:mu-sc-incl-3}
			\bigl\{&(x_1,\zeta_1; \ldots; x_n,\zeta_n) \in (\ctsp{\bb{R}^4})^{\times n} \setminus \cal{Z}^{\times n} : \notag \\
			&\qquad \qquad (0,\ldots,0, \zeta_1,\ldots,\zeta_n, 0,\ldots,0) \in \pi_2(\WF{\omega_{s+n+t}}) \bigr\} 
			\subseteq \Gamma_n,
		\end{align}
		for every $s,t \in \{0,\ldots,N\}$. Altogether, the inclusions \eqref{eq:mu-sc-incl-1}, \eqref{eq:mu-sc-incl-2}, and \eqref{eq:mu-sc-incl-3} prove the fulfillment of the $\mu$SC for the admissible vector state $\omega^\Psi$.
	\end{prf}
	\vsp
	Since the wavefront set of a sum of distributions is contained within the union of their individual wavefront sets, a finite convex combination of admissible vector states also fulfills the $\mu$SC under the criterion of Theorem \ref{thm:mu-sc}.

    \section{Conclusion and Outlook}
    \label{sec:conclusion}

    To recapitulate, an admissible vector state on the warped algebra $\scr{P}(\bb{R}^4, \Sigma_Q)$ fulfills the $\mu$SC if the state $\omega$ on the original algebra $\scr{P}(\bb{R}^4)$ fulfills the $\mu$SC. Recall that $\omega$ refers to the state on the Borchers-Uhlmann algebra $\scr{B}(\bb{R}^4)$ that constructs the GNS Hilbert space in which it is then viewed as a vector state on $\scr{P}(\bb{R}^4)$. On restoring the spectrum condition, as is reasonable in Minkowski spacetime, $\omega$ naturally fulfills the $\mu$SC \cite[Thm.~4.6]{Brunetti1}. Hence, so does every admissible vector state on $\scr{P}(\bb{R}^4, \Sigma_Q)$.
    \par
    Similar to the warped convolution, the algebra $\scr{P}(\bb{R}^4)$ can be deformed with the Rieffel product (\cite{Buchholz2}, \cite{Rieffel}). However, admissible vector states are generally not positive, and must be composed with some operation such as the warped convolution \cite[Lem.~2.4]{Buchholz2} or a (modified) Gaußian convolution \cite{Kaschek} to enforce positivity. In both cases, identical results on the fulfillment of the microlocal spectrum condition can be derived, based on almost identical proofs.
    \par
    The analysis of singular field correlations can be further refined by using Sobolev wavefront sets \cite[App.~B]{Junker} instead of the smooth wavefront set. This corresponds to the generalisation from Hadamard states to adiabatic states \cite{Junker}. It perhaps has the potential to distinguish between $\scr{P}(\bb{R}^4)$ and $\scr{P}(\bb{R}^4, \Sigma_Q)$, since the monomials in \eqref{eq:mu-sc-ineq} suggest that the microlocal Sobolev regularity of the symbolic distribution of a warped $n$-point distribution, and hence of the warped $n$-point distribution itself (with a suitable adaptation of Theorem \ref{thm:osc-int-wfs}), will be lower than that of the corresponding $n$-point distribution. In other words, an adiabatic state on $\scr{P}(\bb{R}^4)$ and $\scr{P}(\bb{R}^4, \Sigma_Q)$ is expected to be of lower order on the latter. One hopes that this distinction relates to the degree of nonlocality in noncommutative Minkowski spacetime.

	\phantomsection
	\addcontentsline{toc}{section}{References}
	\small
	\let\textbf\relax
	\bibliographystyle{amsalpha}
	\bibliography{refs}

\end{document}